\documentclass[
aps, prb,showpacs,superscriptaddress,numerical,amsmath,amssymb,floatfix,
reprint,
]{revtex4-2}
\usepackage[
pdffitwindow=true,
colorlinks=true,
frenchlinks=false,
linkcolor=blue,
anchorcolor=blue,
citecolor=blue,
filecolor=blue,
urlcolor=blue,
bookmarks=true,
bookmarksopen=true,
bookmarksnumbered=true,
bookmarksopenlevel=1,
plainpages=false,
pdfpagelayout=TwoPageLeft,
pdfpagelabels=true,
breaklinks
]{hyperref}
\usepackage[per-mode=symbol,separate-uncertainty]{siunitx}
\usepackage{graphicx}
\usepackage{dcolumn}
\usepackage{url}
\usepackage{color}
\usepackage{graphicx,wrapfig,lipsum}
\usepackage{bm}
\usepackage[english]{babel}
\usepackage{color}
\usepackage{ulem}
\usepackage{float}
\usepackage{lipsum}
\bibpunct{[}{]}{,}{n}{}{}
\usepackage{chemformula}

\begin{document}
\title{Reduced effective magnetization and damping by slowly-relaxing impurities in strained $\gamma$-\ch{Fe2O3} thin films }

\author{M.~M\"uller}
\email{manuel.mueller@wmi.badw.de}
\affiliation{Walther-Mei{\ss}ner-Institut, Bayerische Akademie der Wissenschaften, 85748 Garching, Germany}
\affiliation{Physik-Department, Technische Universit\"{a}t M\"{u}nchen, 85748 Garching, Germany}
\author{M.~Scheufele}
\affiliation{Walther-Mei{\ss}ner-Institut, Bayerische Akademie der Wissenschaften, 85748 Garching, Germany}
\affiliation{Physik-Department, Technische Universit\"{a}t M\"{u}nchen, 85748 Garching, Germany}
\author{J.~G\"uckelhorn}
\affiliation{Walther-Mei{\ss}ner-Institut, Bayerische Akademie der Wissenschaften, 85748 Garching, Germany}
\affiliation{Physik-Department, Technische Universit\"{a}t M\"{u}nchen, 85748 Garching, Germany}
\author{L.~Flacke}
\affiliation{Walther-Mei{\ss}ner-Institut, Bayerische Akademie der Wissenschaften, 85748 Garching, Germany}
\affiliation{Physik-Department, Technische Universit\"{a}t M\"{u}nchen, 85748 Garching, Germany}

\author{M.~Weiler}
\affiliation{Fachbereich Physik and Landesforschungszentrum OPTIMAS,
	Technische Universit\"{a}t Kaiserslautern, 67663 Kaiserslautern, Germany}
\author{H.~Huebl}
\affiliation{Walther-Mei{\ss}ner-Institut, Bayerische Akademie der Wissenschaften, 85748 Garching, Germany}
\affiliation{Physik-Department, Technische Universit\"{a}t M\"{u}nchen, 85748 Garching, Germany}
\affiliation{Munich Center for Quantum Science and Technology (MCQST), 80799 M\"{u}nchen, Germany}
\author{S.~Gepr\"ags}
\affiliation{Walther-Mei{\ss}ner-Institut, Bayerische Akademie der Wissenschaften, 85748 Garching, Germany}
\author{R.~Gross}
\affiliation{Walther-Mei{\ss}ner-Institut, Bayerische Akademie der Wissenschaften, 85748 Garching, Germany}
\affiliation{Physik-Department, Technische Universit\"{a}t M\"{u}nchen, 85748 Garching, Germany}
\affiliation{Munich Center for Quantum Science and Technology (MCQST), 80799 M\"{u}nchen, Germany}
\author{M.~Althammer}
\email[]{matthias.althammer@wmi.badw.de}
\affiliation{Walther-Mei{\ss}ner-Institut, Bayerische Akademie der Wissenschaften, 85748 Garching, Germany}
\affiliation{Physik-Department, Technische Universit\"{a}t M\"{u}nchen, 85748 Garching, Germany}

\date{\today}

\begin{abstract}
We study the static and dynamic magnetic properties of epitaxially strained $\gamma$-\ch{Fe2O3} (maghemite) thin films grown via pulsed-laser deposition on MgO substrates by SQUID magnetometry and cryogenic broadband ferromagnetic resonance experiments. SQUID magnetometry measurements reveal hysteretic magnetization curves for magnetic fields applied both in- and out of the sample plane. From the magnetization dynamics of our thin films, we find a small negative effective magnetization in agreement with a strain induced perpendicular magnetic anisotropy. Moreover, we observe a non-linear evolution of the ferromagnetic resonance-linewidth as function of the microwave frequency and explain this finding with a model based on slowly relaxing impurities, the so-called slow relaxor model. By investigating the magnetization dynamics in our maghemite thin films as a function of frequency and temperature, we can isolate the temperature dependent contribution of the slowly relaxing impurities  to the resonance linewidth and, in particular, observe a sign change in the effective magnetization. This finding provides evidence for a transition of the magnetic anisotropy from a perpendicular easy axis to an easy in-plane anisotropy for reduced temperatures. 
\end{abstract}
\maketitle

\section{Introduction}
The field of spintronics aims to exploit the electron spin in magnetically ordered materials for data storage and data processing applications. As information is encoded in the angular momentum degree of freedom, charge transport, and therefore the use of electrically conducting materials, is not absolutely required for spintronic devices. In general, magnetic insulators, where spin
information is transported by quantized excitations of the spin system, called magnons, are promising alternatives for the implementation of logic circuits based on angular momentum transport. A particular advantage of a spin current based logic may be the reduction of resistive losses present in charge current based devices \cite{Brataas2020a}. Magnetic insulators are hence of key importance in the emerging fields of magnonics \cite{Chumak2015,Chumak2017}, spin-caloritronics \cite{Bauer2012}, and spin-orbitronics \cite{Li2016}. While there is a large variety of magnetically ordered insulators, most of them have not yet been explored in depth regarding their magnetic properties. At present, the majority of studies in the field is focused on yttrium iron garnet (\ch{Y3Fe5O12}, YIG), as its low Gilbert damping $\alpha\simeq 10^{-5}$ \cite{Cherepanov1993,Spencer1959, Maier-Flaig2017, Klingler2017} is one of the key parameters for spin-wave based devices. In particular, the low damping allows for a spin wave propagation length up to the millimeter regime at frequencies ranging from the GHz to the THz regime \cite{Serga2010,Seifert2018}. However, apart from its nice damping properties, YIG has some drawbacks regarding the integration into more complex heterogeneous devices. In this respect, the room-temperature ferrimagnetic insulator $\gamma$-\ch{Fe2O3} (maghemite) is more promising, offering application perspectives for example as magnetic tunnel barrier for spin-filter devices \cite{Grau-Crespo2010}, magnetic recording media \cite{Huang2013, Dronskowski2001}, as well as for microwave- \cite{Palfalvi2005} and thermoelectric devices \cite{Jimenez-Cavero2017}. 

This motivates us to study maghemite from a materials perspective. To this end,  we optimize the epitaxial growth of maghemite thin films and characterize the static and dynamic magnetic properties of the iron oxide-based ferrimagnetic insulator $\gamma$-\ch{Fe2O3}. However, the growth of high quality $\gamma$-\ch{Fe2O3} thin films is challenging as this phase is metastable and tends to easily transform into antiferromagnetic hematite ($\alpha$-\ch{Fe2O3}), which is the equilibrium phase above $350^\circ $C \cite{Lee2001,Dghoughi2006}. Hence, to obtain pure $\gamma$-\ch{Fe2O3}, an established method is the deposition on substrate materials with a slight lattice mismatch, such as MgO \cite{Alraddadi2020, Jimenez-Cavero2017} or \ch{Al2O3} \cite{Huang2013}. Here, epitaxially strained maghemite thin films are grown via pulsed laser deposition (PLD) on MgO (001) substrates. The good crystalline quality and coherently strained growth of our films is verified by high-resolution X-ray diffraction (XRD). SQUID magnetometry is performed to study the static magnetic properties of the thin films. The thereby recorded magnetization curves provide evidence for a finite out-of-plane (oop) magnetic anisotropy contribution. Furthermore, the dynamic magnetic properties of our thin film samples are investigated via broadband ferromagnetic resonance (bbFMR).
In detail, we find a small negative effective magnetization $M_{\mathrm{eff}}$, suggesting the presence of a strain-induced perpendicular magnetic anisotropy in our samples. This makes $\gamma$-\ch{Fe2O3} particularly desirable for all-electrical magnon transport experiments \cite{Guckelhorn2021, Divinskiy2021, Evelt2018, Brataas2020a} as it lowers non-viscous contributions to damping and thereby increases the effective magnon conductivity. For the temperature dependence of the effective magnetization, $M_{\mathrm{eff}}(T)$, we observe a sign change on reducing temperature. This indicates a transition of the magnetic anisotropy in our maghemite thin films from an out-of-plane (oop) easy-axis to an in-plane (ip) easy-plane orientation. Furthermore, we also observe a non-linear behavior in the FMR linewidth $\Delta H(f)$ as a function of the microwave frequency $f$. This behavior can be well interpreted in terms of a model based on slowly relaxing impurities \cite{VanVleck1963, Nembach2011, Chen1990, Woltersdorf2009, Maier-Flaig2017}. In our experiments, we studied two maghemite thin films of different thickness ($45.0$\,nm and $52.6$\,nm) but comparable static and dynamic magnetic 
properties. Since the thicker film allowed for a better signal-to-noise ratio in the cryogenic bbFMR experiments, we only present the dynamic magnetic properties of the thicker film in the main text and shift those of the thinner film into the Appendix \ref{App: V}.

This article is organized as follows:
In Section~\ref{Ch: growth} we describe the growth and structural characterization of our thin films. In Section~\ref{Ch: magdym} we then focus on the static and dynamic magnetic properties of our $\gamma$-\ch{Fe2O3} films before discussing the origin of the slowly-relaxing impurity damping as well as the presence of additional non-linear damping mechanisms in Section~\ref{Section: discussion}. Finally, in Section~\ref{Ch: Conclude} we summarize our key findings.

\section{Thin film deposition and structural characterization}
\label{Ch: growth}
The investigated $\gamma$-\ch{Fe2O3} maghemite films are grown via pulsed laser deposition on MgO (001) substrates using a substrate temperature of $T_{\mathrm{S}}=320\,^\circ$C, an oxygen pressure of $p_{\mathrm{O_2}}=25\,\mu$bar, a laser fluence at the stochiometric, polycrystalline $\alpha$-\ch{Fe2O3} target of $\rho_{\mathrm{L}}=2.5 \mathrm{{\,J}/{cm^2}}$ and a laser pulse repetition frequency of $f=2$ Hz. Due to the lattice mismatch between the substrate and the thin film, the pseudomorphically grown $\gamma$-\ch{Fe2O3} thin films exhibit tensile strain within the film plane. The $\gamma$-\ch{Fe2O3} unit cell is found to grow in a cubic phase on four MgO unit cells. This results in a lattice mismatch of $\epsilon=(2a_{\mathrm{MgO}}^{\mathrm{bulk}}-a_{\mathrm{\gamma-\ch{Fe2O3}}}^{\mathrm{bulk}})/2a_{\mathrm{\ch{MgO}}}^{\mathrm{bulk}}=1.1\,\%$ using the bulk lattice constants $a_{\mathrm{\gamma-\ch{Fe2O3}}}^{\mathrm{bulk}}=0.8332$\,nm \cite{Jorgensen2007} and $a_{\mathrm{\ch{MgO}}}^{\mathrm{bulk}}=0.4212$\,nm \cite{Villars2016:sm_isp_sd_0310770}. 

\begin{figure}[tbh]	
	\centering
	\includegraphics[width=1.0\columnwidth, clip]{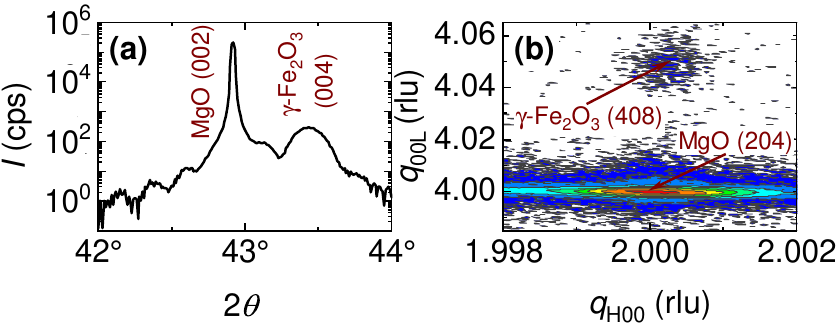}
	\caption{(a) $2\theta-\omega$ XRD scan of the 45.0 nm thin $\gamma$-\ch{Fe2O3} sample. Laue oscillations of the $\gamma$-\ch{Fe2O3} (004)-reflection are visible. (b) Reciprocal space mapping around the asymmetrical MgO (204) and $\gamma$-\ch{Fe2O3} (408) reflection. The units are given in reciprocal lattice units (rlu) with respect to the MgO substrate. From the thin film reflection at $q_{\mathrm{H00}}=1.99998$\,rlu and $q_{\mathrm{00L}}=4.0463$\,rlu, we determine an epitaxial strain of $\epsilon_{\mathrm{xx}} =1.1\%$ in the film plane and $\epsilon_{\mathrm{zz}} =-0.05\%$ out of the film plane.}
	\label{Fig: 1}
\end{figure}

The crystalline quality as well as the epitaxial strain of the maghemite thin films are analyzed by XRD. Fig.\,\ref{Fig: 1}(a) depicts a $2\theta$-$\omega$ scan showing the XRD intensity $I$ in the vicinity of the MgO (002) reflection. At $2\theta=43.44^\circ$, we observe the $\gamma$-\ch{Fe2O3} (004) diffraction peak, which is broadened due to the finite film thickness of 45.0 nm. In addition, we observe Laue oscillations indicating a coherent growth of $\gamma$-\ch{Fe2O3} on MgO (001). From the $2\theta$ position of the $\gamma$-\ch{Fe2O3} (004) reflection, we calculate the oop lattice constant to $c=0.8327$\,nm. Additionally, to determine the ip lattice constant $a_{\mathrm{\gamma-\ch{Fe2O3}}}$ of maghemite, a reciprocal space mapping around the maghemite (408) reflection is performed and shown in Fig.\,\ref{Fig: 1}(b). We observe the (408) reflection of $\gamma$-\ch{Fe2O3} at $q_{\mathrm{H00}} =1.99998$\,reciprocal lattice units (rlu) and $q_{\mathrm{00L}}=4.0463$\,rlu, yielding the lattice constants $a_{\mathrm{\gamma-\ch{Fe2O3}}} = 0.8423$ nm and $c_{\mathrm{\gamma-\ch{Fe2O3}}}=0.8326$ nm. The latter is in excellent agreement with the value extracted from the (004) $\gamma$-\ch{Fe2O3}-reflection in panel (a). The in-plane lattice constant results in an epitaxial strain in the thin film plane of $\epsilon_{\mathrm{xx}} =1.1\%$ indicating a fully epitaxially strained thin film. The out-of-plane strain is determined to $\epsilon_{\mathrm{zz}} =-0.05\%$. This value is significantly lower than that derived under the naive assumption that maghemite exhibits the same Poisson ratio as hematite ($\nu\approx 0.12$) \cite{Chicot2011}, which would lead to $\epsilon_{\mathrm{zz}}=-2\epsilon_{\mathrm{xx}}/(1-\nu)=-2.5\,\%$. We assign the unexpectedly low out-of-plane strain and the correspondingly large increase of the unit cell volume of $\Delta V/V\simeq 1.0\,\%$ to an oxygen deficiency of our maghemite thin films, although the film growth was carried out in a pure oxygen atmosphere. 

\begin{figure}[tbh]	
	\centering
	\includegraphics[width=1.0\columnwidth, clip]{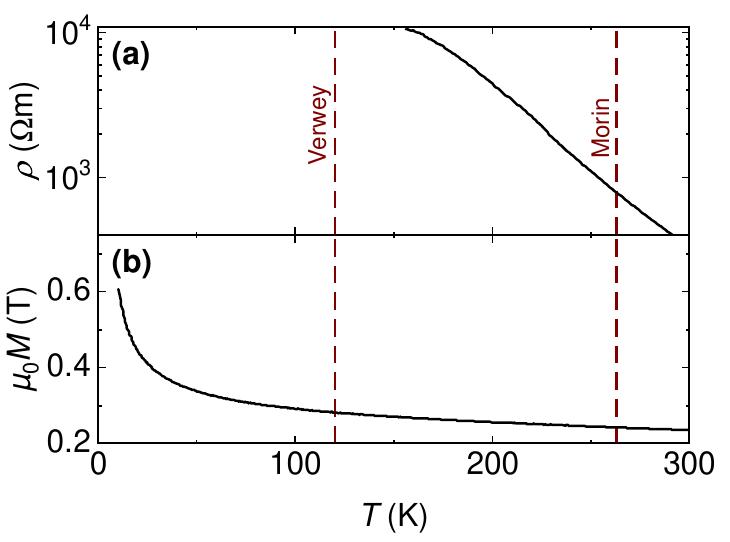}
\caption{(a) Temperature dependence of the electrical resistivity $\rho$ of a 45.0\,nm thick $\gamma$-\ch{Fe2O3} sample. Below $T=155$\,K, the resistance of the investigated film exceeded the range of our measurement scheme. Hence, below $T=155$\,K, $\rho$ is above the detection limit. (b) Temperature dependence of the magnetization $M$ of the same film measured at an external field of $\mu_0H_{\mathrm{ext}}=0.5$\,T. The film was previously cooled in a magnetic field of 7 T to saturate its magnetization. The brown dashed lines mark the temperatures of the Morin and Verwey transition in $\alpha$-\ch{Fe2O3} and \ch{Fe3O4}, respectively. }
	\label{Fig: SI-Resistivity}
\end{figure}

To demonstrate the growth of a pure $\gamma$-\ch{Fe2O3} phase, we measure the temperature dependence of the electrical resistivity in a 4-probe configuration using the Van-der-Pauw-method \cite{VanderPAUW1991a}. The result is shown in Fig.\,\ref{Fig: SI-Resistivity}(a), while in (b) we plot the temperature dependence of the magnetization $M$ measured at a magnetic field $\mu_0H_{\mathrm{ext}}=0.5$\,T applied within the plane of the 45.0\,nm thick $\gamma$-\ch{Fe2O3} thin film. Fig.\,\ref{Fig: SI-Resistivity} clearly demonstrates that both $\rho(T)$ and $M(T)$ show a smooth temperature dependence and, hence, no evidence for the presence of undesired iron oxide impurity phases such as hematite ($\alpha$-\ch{Fe2O3}) and magnetite (\ch{Fe3O4}). The latter are expected to result in abrupt changes in both $\rho(T)$ and $M(T)$ at the temperatures of the Morin transition \cite{Morin1950} ($T=263$\,K) or the Verwey transition (\ch{Fe3O4}) \cite{VERWEY1939} ($T=120$\,K), respectively. 

In summary, the study of the structural properties of our $\gamma$-\ch{Fe2O3} films shows that we are able to grow single-phase $\gamma$-\ch{Fe2O3} thin films of highly crystalline quality but most likely with a finite density of oxygen vacancies. The latter are relevant for the magnetic properties of the films which are discussed in the subsequent section~\ref{Ch: magdym}.

\section{Magnetic properties of \texorpdfstring{$\boldsymbol{\gamma-\mathrm{Fe_2O_3}}$}{blank} maghemite films}
\label{Ch: magdym}
\subsection{Static magnetic properties}
\begin{figure}[tbh]	
	\centering
	\includegraphics[width=1.0\columnwidth, clip]{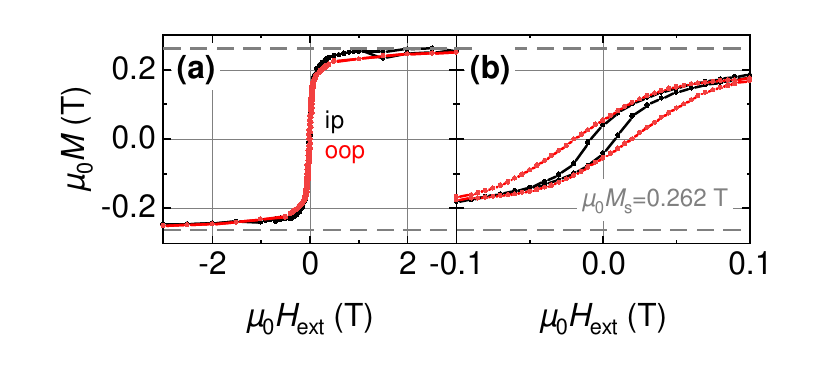}
	\caption{(a) Room-temperature magnetization versus applied magnetic field curves obtained by SQUID magnetometry for $\gamma$-\ch{Fe2O3} thin films in ip (black) and oop (red) geometry. (b) Expanded view of the magnetic hysteresis curve shown in (a) for the field range $-0.1\mathrm{\,T}\leq\mu_0H_{\mathrm{ext}}\leq0.1\mathrm{\,T}$. The horizontal, gray dashed lines mark the saturation magnetization of $\mu_0M_{\mathrm{s}}=0.262$\,T. }
	\label{Fig: 3}
\end{figure}

For magnetic characterization we measure room-temperature magnetic hysteresis curves via SQUID magnetometry with the external magnetic field $H_{\mathrm{ext}}$ applied both within the thin film plane (ip) and perpendicular to it (oop). The hysteresis curves of the 45.0\,nm thick maghemite sample are shown in Fig.\,\ref{Fig: 3} for the ip (black circles) and oop (red circles) geometry. Here, we substract a diamagnetic linear background contribution of the MgO substrate. In panel (a), we show the entire examined field-range from $-3\mathrm{\,T}\leq\mu_0H_{\mathrm{ext}}\leq3\mathrm{\,T}$, while panel (b) displays the field range $-0.1\mathrm{\,T}\leq\mu_0H_{\mathrm{ext}}\leq0.1\mathrm{\,T}$. We observe a hysteretic behavior in $M(H)$ in both ip and oop geometry. This is a clear hint to the presence of an extra magnetic anisotropy contribution in addition to the ip shape anisotropy of our thin films. We attribute this magnetic anisotropy contribution to the strain present in the maghemite film in combination with the magnetoelastic coupling \cite{Popova2001}. 

For the saturation magnetization, we extract $\mu_0M_{\mathrm{s}}=0.262$\,T, which is only about half of the bulk value $\mu_0M_\mathrm{s}\simeq0.5$\,T \cite{Huang2013, Jimenez-Cavero2017}. However, according to literature, the saturation magnetization of $\gamma$-\ch{Fe2O3} thin films is known to depend on the fabrication method and the thin film quality. For example, reduced values of $M_{\mathrm{s}}$ have been reported for $\gamma$-\ch{Fe2O3} thin films grown on MgO substrate via molecular beam epitaxy ($\mu_0M_{\mathrm{s}}=0.339$\,T) \cite{Alraddadi2020}. The most likely origin of the strong reduction of $M_\mathrm{s}$ in our thin films grown on MgO compared to the bulk value is the presence of a finite density of so-called antiphase boundaries (APBs) \cite{Alraddadi2020,Bobo2001, Singh2017}. The APBs are formed when crystalline regions of maghemite with different symmetry merge and couple antiferromagnetically during film growth \cite{Lubitz2001,McMichael2000}. We note that APBs are commonly observed in iron oxides and cause domain wall pinning and consequently require a large external magnetic field to align the orientation of all of the individual magnetic domains along the magnetic field direction (cf. Ref.\,\cite{Knittel2006}). In our SQUID magnetometry measurements their contribution is difficult to quantify, since it is masked by the diamagnetic contribution from the MgO substrate at large $H_{\mathrm{ext}}$.

\subsection{Dynamic magnetic properties}

\begin{figure}[tbh]	
	\centering
	\includegraphics[width=1.0\columnwidth, clip]{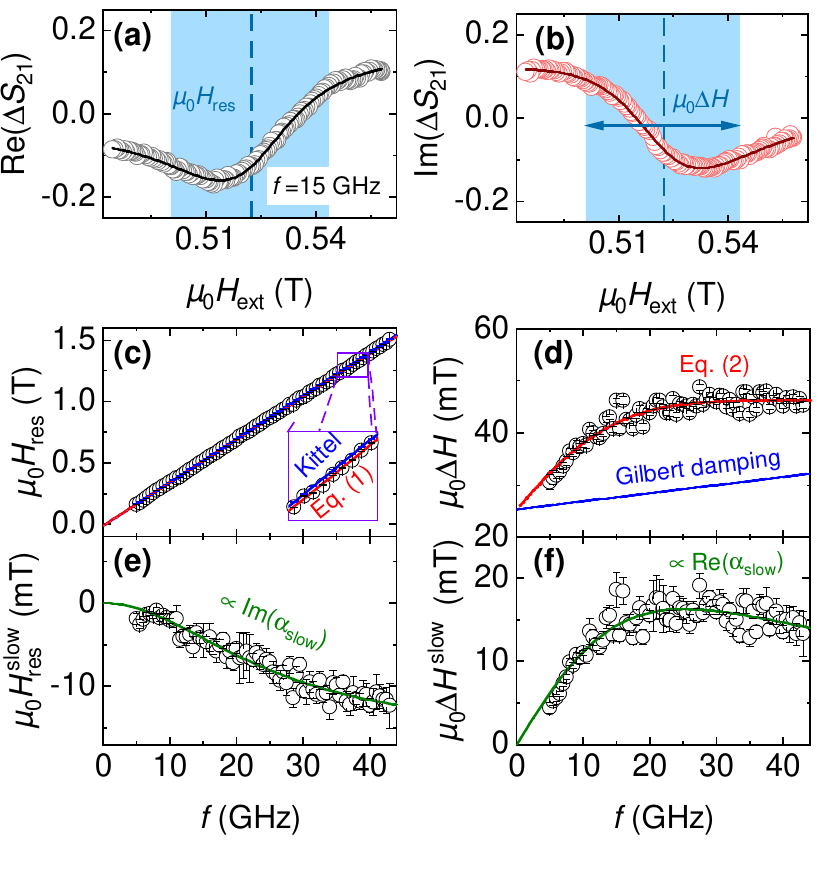}
	\caption{Room temperature bbFMR data of the 52.6\,nm thick $\gamma$-\ch{Fe2O3} thin film taken with the external magnetic field applied out-of-plane. Exemplary raw data for (a) the real and (b) the imaginary part of the complex microwave transmission parameter $S_{21}$ as function of the applied magnetic field $H_{\mathrm{ext}}$ recorded at a microwave frequency of $f=15$\,GHz. The continuous lines represent fits to the Polder susceptibility. The dashed line and colored box represent the resonance field $H_{\mathrm{res}}$ and linewidth $\Delta H$, respectively. (c) Raw data of the extracted resonance field $H_{\mathrm{res}}$ together with a fit following Eq.\,(\ref{Hres}) (red line). The inset shows the small difference between the linear Kittel contribution (blue line) and the total fit (red line). (d) Raw data of the extracted resonance linewidth $\Delta H$ together with a fit following Eq.\,(\ref{DeltaH}) (red line). The linear-in-frequency term in Eq.\,(\ref{DeltaH}) is the Gilbert damping (blue line). (e) Slowly relaxing impurity contribution to the resonance field $H_{\mathrm{res}}^{\mathrm{slow}}$ obtained by subtracting the Kittel contribution (blue line) from the raw data for $H_{\mathrm{res}}$ in panel (c) together with a fit to $\mathrm{Im}(\alpha_{\mathrm{slow}})$ according to Eq.\,(\ref{Eq Ima}) (green line). (f) Slowly relaxing impurity contribution to the resonance linewidth $\Delta H^{\mathrm{slow}}(f)$ obtained by subtracting the linear Gilbert contribution (blue line) from the raw data for $\Delta H$ in panel (d) together with a fit to $\mathrm{Re}(\alpha_{\mathrm{slow}})(f)$ according to Eq.\,(\ref{Eq Ima}) (green line). }
	\label{Fig: 4}
\end{figure}

To determine the dynamic magnetic properties of our maghemite samples, we perform broadband ferromagnetic resonance (bbFMR) experiments. In particular, we record the complex microwave transmission parameter $S_{21}$ for fixed microwave frequencies in the range $5 \mathrm{\,GHz}\leq f\leq 43.5 \mathrm{\,GHz}$ as a function of the static applied magnetic field $H_{\mathrm{ext}}$ using a vector network analyzer (VNA). We obtain the net change of the complex transmission parameter $\Delta S_{21}$ by $
\Delta S_{21}=(S_{21}-S_{21}^0)/S_{21}$, where $S_{21}^0$ is the value of the off-resonant transmission background. The external static magnetic field is applied along the oop direction to suppress two-magnon scattering \cite{Hurben1998}. Exemplary raw data for the real (a) and imaginary part (b) of the complex microwave transmission parameter $\Delta S_{21}$ as function of external magnetic field $H_{\mathrm{ext}}$ is shown in Fig.\,\ref{Fig: 4}. The complex $\Delta S_{21}$ data is fitted to the Polder susceptibility $\chi_{\mathrm{P}}$ \cite{Polder1949, Shaw2014} to extract the resonance field $H_{\mathrm{ext}}$ (blue dashed line) and the linewidth $\Delta H$ (light blue box) as a function of $f$. The resulting $H_{\mathrm{res}}(f)$ and $\Delta H(f)$ curves of the $52.6$ nm thick maghemite film are plotted in Fig.\,\ref{Fig: 4}(c) and (d). 

\begin{figure}[tbh]	
	\centering
	\includegraphics[width=1.0\columnwidth, clip]{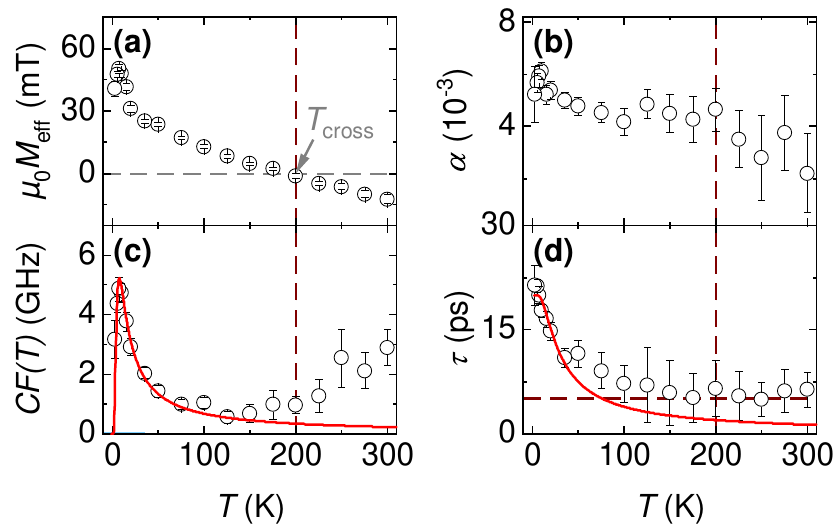}
	\caption{Magnetization dynamics- and slowly relaxing impurity parameters of the $52.6$ nm thick maghemite sample as a function of temperature $T$. (a) Effective magnetization $M_{\mathrm{eff}}$ and (b) Gilbert damping parameter $\alpha$ as function of $T$. (c) Magnitude of slowly relaxing impurity contribution $CF(T)$ together with a theoretical fitting curve for $CF(T)$ (red line) following the product of Eqs.\,(\ref{Eq: C}) and (\ref{Eq: F}). We extract $E_{\mathrm{slow}}=(0.50\pm0.04)$\,meV and $C\cdot T=(66\pm4)$\,GHz$\cdot$K. (d) Relaxation time $\tau$ as function of $T$. The red line represents a fit to Eq.\,(\ref{Eq: time}). We extract $E_{\mathrm{A}}=(3.3\pm0.6)$\,meV and $\tau_0=(19.9\pm0.1)$\,ps.}
	\label{Fig: 5}
\end{figure}

For the FMR linewidth plotted in Fig.\,\ref{Fig: 4}(d), we observe a non-linear $\Delta H(f)$ dependence with a distinct change in slope at intermediate frequencies ($10\mathrm{\,GHz}\leq f \leq 20\mathrm{\,GHz}$). In previous studies\cite{Woltersdorf2009 ,Nembach2011, Chen1990}, such a non-linear behavior of $\Delta H(f)$ has been attributed to the slowly relaxing impurity mechanism. Here, an additional contribution to magnetization damping is induced due to the exchange coupling of the magnetization of the thin film and the electron spin of an impurity atom. The excitation transferred to the electronic spin then relaxes via the emission of a phonon. This transfer naturally depends on the transition frequency of the participating state of the impurity and its thermal occupation, rendering this effect frequency and temperature dependent. The contribution of the slowly relaxing impurities acts in addition to the Gilbert damping, which exhibits a linear relation of $\Delta H$ and $f$. In the presence of slowly relaxing impurities and assuming a uniform uniaxial out-of-plane anisotropy field $H_{\mathrm{k}}$ parallel to the applied static magnetic field, the total dispersion of the ferromagnetic resonance and the change of linewidth with frequency can then be described by \cite{Nembach2011, Woltersdorf2009}
\begin{align}
	\mu_0 H_{\mathrm{res}}&=\mu_0 M_{\mathrm{eff}}+\frac{h f}{g\mu_{\mathrm{B}}}[1+\mathrm{Im}(\alpha_{\mathrm{slow}}(f))]
	\label{Hres}
\\
	\mu_\mathrm{0}\Delta H&=\mu_\mathrm{0}H_{\mathrm{inh}}+2\cdot\frac{h f}{g\mu_{\mathrm{B}}}[\alpha+\mathrm{Re}(\alpha_{\mathrm{slow}}(f))].
	\label{DeltaH}
\end{align}
Here, $\mu_0$ is the vacuum magnetic permeability, $\mu_\mathrm{B}$ the Bohr magneton, and $h$ the Planck constant. Furthermore, the parameters describing the magnetization dynamics are the $g$-factor $g$, the effective magnetization $M_{\mathrm{eff}}=M_{\mathrm{s}}-H_{\mathrm{k}}$, the Gilbert damping parameter $\alpha$ and the inhomogeneous linewidth broadening $H_{\mathrm{inh}}$. Finally, we account for the slow-relaxor mechanism via \cite{Nembach2011}
\begin{equation}
\alpha_{\mathrm{slow}}(f)=CF(T)\left[\frac{\tau}{1+(2\pi f\tau)^2}-i\frac{2\pi f\tau^2}{1+(2\pi f\tau)^2}\right],
\label{Eq Ima}
\end{equation}
with the relaxation time $\tau$ of the slowly relaxing impurities. The constant $C$ is given by \cite{Woltersdorf2009, Nembach2011}
\begin{equation}
C=\frac{g\mu_{\mathrm{B}}N_{\mathrm{slow}}}{8 M_{\mathrm{s}}\hbar k_{\mathrm{B}}T}\left[\left(\frac{\partial E_{\mathrm{slow}}}{\partial\phi}\right)^2+\left(\frac{\partial E_{\mathrm{slow}}}{\partial\theta}\right)^2\right].
\label{Eq: C}
\end{equation}
Here, $N_{\mathrm{slow}}$ is the concentration of the slowly relaxing impurities and $E_{\mathrm{slow}}$ is the spin splitting of the paramagnetic impurity energy levels induced by the exchange field of the magnetically ordered layer. The angular derivatives of $E_{\mathrm{slow}}$ (where $\theta$ and $\phi$ are the azimuthal and polar angles with respect to the magnetic field axis, respectively) are a consequence of the anisotropic exchange coupling of the impurity atoms with the ferromagnetic moment \cite{Nembach2011}.
The function $F(T)$ describes the temperature dependence of the damping contribution by the slowly relaxing impurities and can be expressed as 
\begin{equation}
	F(T)=\mathrm{sech}^2\left(\frac{E_{\mathrm{slow}}}{k_{\mathrm{B}} T}\right).
	\label{Eq: F}
\end{equation}
The red lines in Fig.\,\ref{Fig: 4}(c) and (d) represent fits to Eqs.\,(\ref{Hres}) and (\ref{DeltaH}) including the parameters describing the impact of the slowly relaxing impurities, while the blue lines represent the behavior without the slow-relaxor mechanism, extracted from the fits. Figures~\ref{Fig: 4}(e) and (f) show the isolated slow-relaxor contribution to  the FMR resonance field and linewidth as function of the excitation frequency. The green lines indicate the real and imaginary part of $\alpha_{\mathrm{slow}}$ following Eq.\,(\ref{Eq Ima}). While the imaginary part of $\alpha_{\mathrm{slow}}(f)$ in Fig.\,\ref{Fig: 4}(c) modifies the FMR resonance frequency only weakly and could be (mis-)interpreted in terms of a modified $g$-factor, the real part of $\alpha_{\mathrm{slow}}(f)$ shown in Fig.\,\ref{Fig: 4}(d) well describes the non-linear variation of the FMR linewidth. Hence, we apply a global fit for both $H_{\mathrm{res}}(f)$ and $\Delta H(f)$ using the same shared parameters to extract the contribution due to slowly relaxing impurities to $H_{\mathrm{res}}$. We also use a temperature-independent $g$-factor $g=2.022$, which matches previous results for maghemite nanopowders \cite{Youssef2006}, to reduce the number of free fitting parameters. The fitted magnetization dynamics and parameters describing the slowly relaxing impurities represent the room-temperature data points in Fig.\,\ref{Fig: 5}. We note that the extrapolated negative $H_{\mathrm{res}}$ value at $f=0$ translates into $M_{\mathrm{eff}}<0$. This clearly indicates a dominant oop easy-axis anisotropy with $\mu_0H_{\mathrm{k}}\approx0.3$ T for our maghemite thin films.

Figure~\ref{Fig: 5} shows the parameters extracted by fitting the data of our bbFMR experiments on the $52.6$\,nm thick maghemite thin film [(a)-(d)] including the slowly relaxing impurity parameters [(c) and (d)] as a function of temperature $T$. We observe a small negative effective magnetization $\mu_0M_{\mathrm{eff}}\approx -12$\,mT at room temperature [see panel (a)], which gradually increases with decreasing temperature. $M_{\mathrm{eff}}$ changes sign at at $T_{\mathrm{cross}}\approx 200$\,K. We attribute the change in $M_{\mathrm{eff}}$ in parts to a reduction in the strain-induced anisotropy contribution $H_{\mathrm{k}}$ due to the reduced strain between MgO and $\gamma$-\ch{Fe2O3} with decreasing temperature, as iron oxides exhibit a larger thermal expansion coefficient \cite{Takeda2009} compared to that of MgO \cite{Madelung1999}, while simultaneously the saturation magnetization $M_{\mathrm{s}}$ of maghemite \cite{Overview2015} increases. In Fig.\,\ref{Fig: 5}(b), we observe an increased Gilbert-damping parameter $\alpha$ with decreasing temperatures. Regarding the characteristic parameters of the slowly relaxing impurities in Fig.\,\ref{Fig: 5}(c), we observe a maximum in $CF(T)$ at $T\simeq 7$\,K followed by a decrease up to 150\,K. Then the magnitude of $CF(T)$ increases again up to room temperature. The peak-behavior at low $T$ is characteristic of damping effects due to slowly relaxing impurities and we observe good agreement of our experimental data with a fit given by the product of Eqs. (\ref{Eq: C}) and (\ref{Eq: F}) to our data for $T\leq50$\,K. We obtain the fitting parameters $E_{\mathrm{slow}}=(0.50\pm0.04)$\,meV and $C\cdot T=(66\pm4)$\,GHz$\cdot$K, which are comparable to previous results for slowly relaxing impurities in literature \cite{Drovosekov2020, Nembach2011}. A potential mechanism to explain the increase in $CF(T)$ for $T>150$\,K is discussed in Section~\ref{Section: discussion}. Finally, for the relaxation time $\tau$ in Fig.\,\ref{Fig: 5}(d), we observe a constant $\tau$ for $T>100$\,K. For temperatures below 100\,K, we fit $\tau$ assuming the relaxation of slowly relaxing impurities via a single particle Orbach process \cite{Bleaney1961,Rfsoyay1961} with activation energy $E_{\mathrm{A}}$
 \begin{equation}
 	\tau=\tau_0\cdot \mathrm{tanh}\left(\frac{E_\mathrm{A}}{2k_{\mathrm{B}}T}\right).
 	\label{Eq: time}
 \end{equation}
We obtain $E_{\mathrm{A}}=(3.3\pm0.6)$\,meV and $\tau_0=(19.9\pm0.1)$\,ps, which correspond well to results in Refs.\,\cite{Drovosekov2020, Hansen1973}. In the Appendix \ref{App: V}, these results are compared to those of the $45$\,nm thick maghemite film. All in all, we observe a comparable magnitude and evolution with temperature for both $CF(T)$ and $\tau$.

\section{Discussion}
\label{Section: discussion}
\subsection{Identification of potential slowly relaxing impurities}
\begin{figure}[tbh]	
	\centering
	\includegraphics[width=1.0\columnwidth, clip]{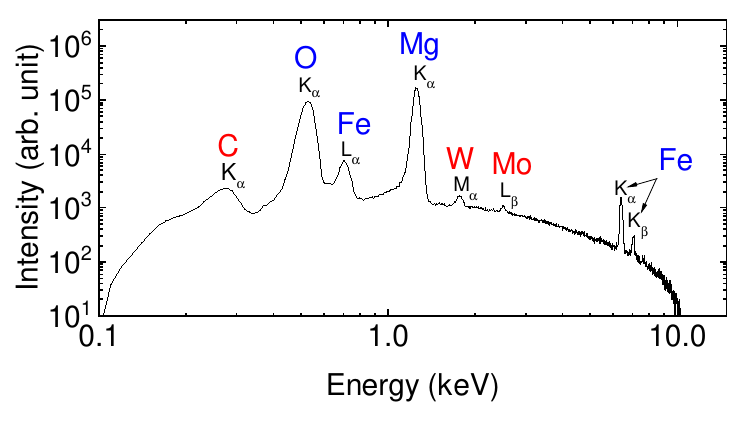}
	\caption{EDS energy spectrum of the 52.6\,nm thick $\gamma$-\ch{Fe2O3} thin film. Visible peaks are labeled with the corresponding element and energy shell. Blue labels correspond to peaks from elements, which comprise the MgO substrate or the $\gamma$-\ch{Fe2O3} thin film, whereas red labels represent peaks from unexpected elements, which we attribute to the sample mount and point source cathode of the SEM. }
	\label{Fig: Raw-data2}
\end{figure}

To identify potential rare-earth impurities in our samples, we perform energy-dispersive X-ray spectroscopy (EDS) on our samples. In Fig.\,\ref{Fig: Raw-data2}, we plot the integrated detected X-ray intensity as a function of X-ray energy $E$ for an area of $50\times40\,\mathrm{\mu m^2}$  of the 52.6\,nm thick $\gamma$-\ch{Fe2O3} thin film. The peaks in the the EDS spectrum allow to identify the atomic species present. The pronounced signatures of oxygen, magnesium and iron atoms is in agreement with the the use of the MgO substrate and the $\gamma$-\ch{Fe2O3} thin film (blue labels). The small $K_\alpha$ peak from carbon is attributed to the  G3347 conductive adhesive tape fixing the sample to the sample holder, whereas the tungsten $M_{\alpha}$-peak most likely stems from the point source cathode of the SEM. The peak at $E\approx 2.504$\,keV can not be directly matched to the energy shell of any element. However, we speculate that it could result from the molybdenum $L_{\mathrm{\beta,2}}$-shell at $E\approx 2.518$\,keV, as the SEM sample stub contains molybdenum. To our knowledge, neither C, Mo nor W impurities have been found to induce a slowly relaxing impurity damping mechanisms. As there are no additional peaks from elements, known to induce slowly relaxing impurity damping such as e.g. Si \cite{Hartwick1968}, Ir \cite{Hansen1973}, Yb \cite{Bleaney1961} and Ge \cite{Nembach2011}, they constitute a mass fraction for our maghemite thin films of less than $w<0.1$\,wt\%, which is the detection limit of the performed EDS experiments. This fraction is expected to be too low to affect the magnetization dynamics in our samples. A remaining plausible candidate are unpaired Fe$^{2+}$-ions \cite{ Spencer1961,Hartwick1968,Hansen1973a,Epstein1967} caused by the oxygen deficiency in our samples (see section \ref{Ch: growth}). It is worth mentioning that a finite concentration of Fe$^{2+}$-ions in $\gamma$-\ch{Fe2O3} nanocrystals has been observed via X-ray absorption spectroscopy in Ref.\,\cite{Coduri2020}.

\subsection{Possible additional damping processes}
It is evident from Fig.\,\ref{Fig: 5} that the observed behavior of the FMR linewidth $\Delta H$ as function of frequency $f$ and temperature $T$ can be well described with the slowly relaxing impurity model. However, there exists a variety of other mechanisms that can cause a non-linear frequency dependence of $\Delta H$. In the following, we list these mechanisms and provide arguments why we identify them as implausible causing the non-linear damping contribution observed for the $\gamma$-\ch{Fe2O3} thin film. Finally, we mention a particular damping process that can explain the deviations from a slow-relaxor impurity like damping behavior observed for our samples at elevated temperatures. 

We begin our discussion with the magnetization damping caused by the scattering of the uniform magnetization mode ($k=0$) with optical phonons according to the Kasuya-LeCraw-mechanism \cite{Kasuya1961, SparksMarshall1964Ft}
or with optical magnons as described by the Kolokolov-L’vov-Cherepanov process \cite{Cherepanov1993}. However, these processes lead to a linewidth variation that is approximately linear in frequency and temperature for intermediate temperatures ($150\mathrm{\,K}\leq T\leq 350\mathrm{\,K}$) \cite{Maier-Flaig2017} and can hence not explain the observed non-linear features in FMR linewidth $\Delta H$ and its temperature dependence shown in Fig.\,\ref{Fig: 5}(c). Similarly, non-Gilbert damping can arise due to two-magnon scattering. However, this process does not exhibit a strong temperature dependence \cite{SparksMarshall1964Ft, gurevich1996magnetization} and is also expected to be suppressed in oop-geometry.

As an alternative explanation for the non-linear frequency dependence of the FMR linewidth observed in our experiments, effects of magnetic anisotropy may play a role. In this context it is worth mentioning that a non-linear $\Delta H(f)$ dependence may be caused by a non-collinear distribution of uniaxial anisotropies with an average angle $\beta$ around the out-of-plane direction. This scenario has been described with the so-called anisotropy dispersion model by Krysztofik \textit{et al.} in Ref.\,\cite{Krysztofik2021}. However, in contrast to our results, they observed a convex shape for the $\Delta H(f)$ dependence with a negative slope in $\Delta H(f)$ such that we can also rule out a distribution in magnetic anisotropy as the origin of the non-linear $\Delta H(f)$ dependence in the high temperature regime.

Finally, we note that there exists another process besides the slowly relaxing impurity mechanism resulting in a peak-like behavior of $\Delta H(T)$, namely the so-called valence-exchange or charge-transfer mechanism \cite{Hartwick1968, Clogston1955, gurevich1996magnetization}. This effect can manifest itself if mixed-valent ions are present in the ferromagnetic sample (i.e. $\mathrm{Fe^{2+}}$ or $\mathrm{Fe^{4+}}$ ions in YIG with nominal only $\mathrm{Fe^{3+}}$-ions). In this case electrons can hop between the different valence lattice sites. This hopping mediates a net energy transfer from the dynamically precessing magnetization to the crystal lattice and can thereby increase the damping. Notably, this mechanism results in the same peak-like frequency dependence to the FMR linewidth \cite{Hartwick1968} $\Delta H_{\mathrm{ct}}\propto \omega \tau_{\mathrm{ct}}/[1+(\omega \tau_{\mathrm{ct}})^2]$ as predicted by the slowly relaxing impurity model [compare Eq.\,(\ref{Eq Ima})]. However, it has a different characteristic timescale $\tau_{\mathrm{ct}}$, representing the electron hopping time. Fortunately, its impact can be distinguished from that of slowly relaxing impurities, as it has been found to manifest itself at temperatures above room temperature, for example above $T\simeq370$\,K in YIG \cite{Hartwick1968, Hansen1973a}. 

\begin{figure}[tbh]	
	\centering
	\includegraphics[width=1.0\columnwidth, clip]{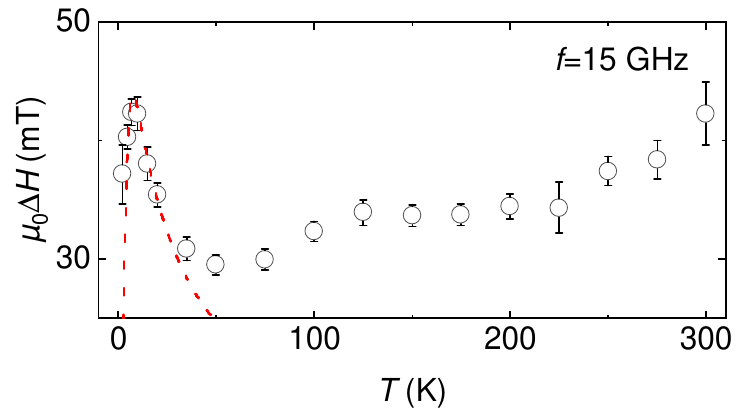}
	\caption{FMR linewidth $\Delta H$ recorded at $f=15$\,GHz for the 52.6\,nm thick maghemite film as function of temperature $T$. The red dashed line represents the slowly relaxing impurity contribution to the FMR linewidth following the real part of Eq.\,(\ref{Eq Ima}) using the parameters extracted from fits to $CF(T)$ and $\tau$ from Fig.\,\ref{Fig: 5}(c) and (d).}
	\label{Fig: Raw-data}
\end{figure}

To discuss the impact of valence-exchange damping in our maghemite thin films, we plot in Fig.\,\ref{Fig: Raw-data} the FMR linewidth $\Delta H$ recorded at $f=15$\,GHz for the 52.6\,nm thick maghemite film as function of temperature $T$. The red dashed line represents the slowly relaxing impurity contribution to the FMR linewidth following the real part of Eq.(\ref{Eq Ima}) using the parameters extracted from fits to $CF(T)$ and $\tau$ from Fig.\,\ref{Fig: 5}(c) and (d). For low temperatures ($T<50$\;K), the peak-like feature in Fig.\,\ref{Fig: Raw-data} can be well described by our slowly relaxing impurity model, while the gradual increase in $\Delta H$ mirrors our observations for the slowly relaxing impurity constant $CF(T)$ in Fig.\,\ref{Fig: 5}(c). This indicates that the increase in $\Delta H$ is generated by another frequency and temperature dependent damping contribution. In this context, the seemingly constant $\tau$ in Fig.\,\ref{Fig: 5}(d) can be attributed to a crossover from the regime dominated by the slowly relaxing impurity mechanism to that dominated by the valence-exchange damping mechanism discussed above. Furthermore, it is worth mentioning that the increase in FMR linewidth at $T>150$\;K coincides with a decrease in resistivity $\rho$ shown in Fig.\,\ref{Fig: SI-Resistivity} and consequently an increased thermally induced hopping probability of the electrons available for the valence-exchange mechanism. Based on these observations, we suspect that the observed increase in FMR linewidth $\Delta H$ and slowly-relaxing impurity constant $CF(T)$ at higher temperatures $T>100$\,K up to room temperature is most likely related to the valence-exchange mechanism caused by electron hopping between $\mathrm{Fe}^{\mathrm{2+}}$ and $\mathrm{Fe}^{\mathrm{3+}}$ ions in maghemite. We do not observe the characteristic high temperature peak for the valence-exchange model, as we performed FMR measurements only up to room temperature in our thin films. Further experiments at temperatures well above room temperature are required to unambiguously identify the responsible mechanism that generates the increase in the magnitude of the peak-like features of the FMR linewidth $\Delta H$ at elevated temperatures.

\section{Conclusion}
\label{Ch: Conclude}
We have performed SQUID magnetometry and temperature dependent bbFMR experiments to characterize the static and dynamic magnetic properties of epitaxially strained $\gamma$-\ch{Fe2O3} thin films grown on MgO substrates. XRD measurements demonstrate a pseudomorphic growth of $\gamma$-\ch{Fe2O3} with a tensile epitaxial strain of $\epsilon_{\mathrm{xx}}=1.1$\% and an unexpectedly small out-of-plane strain of $\epsilon_{\mathrm{zz}}=-0.05$\%, which suggest an oxygen deficiency in our samples. Room temperature SQUID magnetometry reveal hysteretic magnetization curves for the magnetic field applied in both the ip and oop direction. The extracted saturation magnetization $\mu_0M_{\mathrm{s}}=0.262$\,T is only about half of the literature value $\mu_0M_\mathrm{s}\simeq0.5$\,T for bulk material \cite{Huang2013, Jimenez-Cavero2017}. This indicates the presence of a considerable density of antiphase boundaries. Regarding the magnetization dynamics, we find that the Gilbert damping mechanism and an inhomogeneous linewidth broadening is not sufficient to describe the observed frequency dependence of the FMR linewidth, $\Delta H(f)$. We model the apparent cusp-shape in $\Delta H(f)$ by taking slowly relaxing impurities into account. While we cannot precisely pinpoint the impurities responsible for this damping contribution, the presence of a finite density of $\mathrm{Fe}^{2+}$ ions is the most plausible scenario \cite{Spencer1961,Hartwick1968,Hansen1973a,Epstein1967}. This presumption nicely agrees with the observation of an increased unit cell, providing evidence for a significant oxygen deficiency in our samples. The magnetization dynamics- and parameters describing the slowly relaxing impurities are studied as a function of temperature by performing cryogenic bbFMR. For the effective magnetization $M_{\mathrm{eff}}(T)$, we observe a transition of the maghemite from oop easy-axis to ip easy-plane anisotropy induced by the reduced strain in $\gamma$-\ch{Fe2O3} for reduced temperatures. Furthermore, we observe the predicted freeze-out of the slowly relaxing impurity contribution $CF(T)$ at low $T$. For $T>150$\,K, we find a clear increase of the linewidth with increasing temperature. We speculate that this increase is caused by electrons hopping between $\mathrm{Fe}^{2+}$ and $\mathrm{Fe}^{3+}$ ions in our oxygen-deficient films, giving rise to additional damping due to the valence-exchange mechanism. 

Our results can be used as input for the theoretical understanding of damping due to slowly relaxing impurities and might find applications for magnetization damping engineering of magnetic materials for example to enhance the damping for magnetic sensor applications \cite{Nembach2011}. Moreover, the observed out-of-plane anisotropy in maghemite is of interest for energy efficient data storage devices based on magnetic textures such as magnetic bubbles, chiral domain walls, and magnetic skyrmions \cite{Fert2013}. Finally, the strain-induced reduced effective magnetization of maghemite renders it a promising material platform for future magnonics applications of magnetically ordered insulators \cite{Guckelhorn2021, Divinskiy2021, Evelt2018, Brataas2020a}.
\nocite{ VanderPAUW1991a}

\section*{Acknowledgments}
We acknowledge financial support by the Deutsche
Forschungsgemeinschaft (DFG, German Research Foundation) via Germany’s Excellence Strategy EXC-2111-390814868. Furthermore, we want to thank Kedar Honasoge for his help with the EDS X-ray spectroscopy.
\newpage
\appendix
\section{Additional data on magnetic properties of maghemite films}
\label{App: V}
\begin{figure}[H]	
	\centering
	\includegraphics[width=1.0\columnwidth, clip]{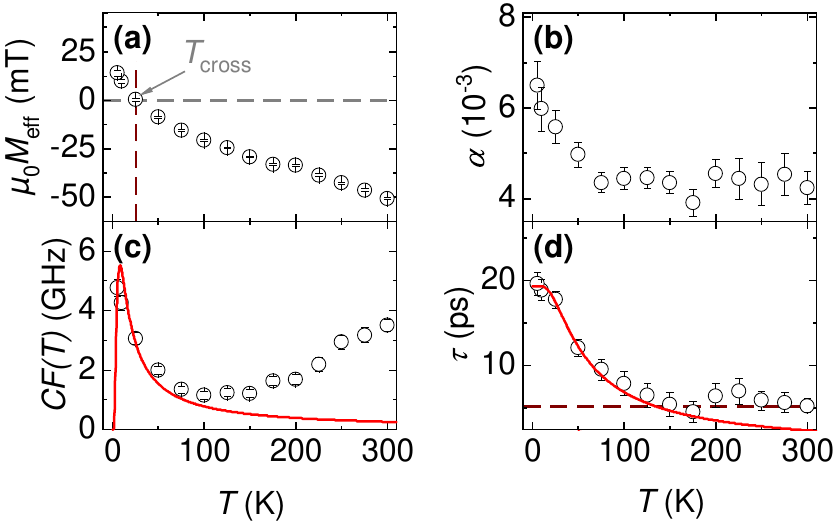}
	\caption{Magnetization dynamics and slowly relaxing impurity parameters of a $45.0$\,nm thick maghemite film as a function of temperature $T$. (a) Effective magnetization $M_{\mathrm{eff}}$ and (b) Gilbert damping parameter $\alpha$ as function of $T$. (c) Magnitude of the slowly relaxing impurity contribution $CF(T)$ together with a theoretical fitting curve for $F(T)$ following the product of Eqs. (\ref{Eq: C}) and (\ref{Eq: F}) (red line). We extract $E_{\mathrm{slow}}=(0.55\pm0.07)$\,meV and $C\cdot T=(78\pm12)$\,GHz$\cdot$K. (d) Relaxation time $\tau$ as a function of $T$. The red line represents a fit to Eq.\,(\ref{Eq: time}) for a single particle Orbach process. We extract $E_{\mathrm{A}}=(6.2\pm0.4)$\,meV and $\tau_0=(19.3\pm0.2)$\,ps. }
	\label{Fig: FeOs2}
\end{figure}

In addition to the data of the $52.6$\,nm thick maghemite thin film we show data on the magnetization dynamics and slowly relaxing impurity parameters of a $45.0$\,nm thick maghemite film as a function of temperature $T$ in Fig.\,\ref{Fig: FeOs2}. The parameters describing the slowly relaxing impurities are again derived under the assumption of a temperature independent $g$-factor $g=2.022$.
The effective magnetization in Fig.\,\ref{Fig: FeOs2}(a) increases continuously with decreasing $T$ and crosses $M_{\mathrm{eff}}=0$ at $T_{\mathrm{cross}}=25$\,K. For the Gilbert damping $\alpha$ in panel (b), we observe an approximately constant $\alpha$ down to $T=25$\,K, followed by a steep increase at lower $T$. Similarly to Fig.\,\ref{Fig: 5}(b), this abrupt increase in damping coincides with the sign change in $M_{\mathrm{eff}}$, indicating that the change in anisotropy in our samples from easy-axis to easy-plane affects the linear damping in our thin films. For the slowly relaxing impurity parameter $CF(T)$ in Fig.\,\ref{Fig: FeOs2}(c), we observe a strong reduction with increasing temperature. The peak-like feature in $CF(T)$ is not evident due to the reduced number of data points in low temperature regime. With increasing temperature, we first observe the predicted reduction in the magnitude of the slowly relaxing impurity contribution, before $CF(T)$ shows an increase for $T>150$\,K in agreement with the results in Fig.\,\ref{Fig: 5}(c) of the main text. We perform a fit to the product of Eqs.\,(\ref{Eq: C}) and (\ref{Eq: F}) (red line) and extract $E_{\mathrm{slow}}=(0.55\pm0.07)$\,meV and $C\cdot T=(78\pm12)$\,GHz$\cdot$K. These values coincide with fitting results for the $52.6$\,nm thick maghemite sample and hence demonstrate that the FMR results are representative for our maghemite films. Finally, for the relaxation time $\tau$ in panel (d), we again observe an approximately constant $\tau$ for elevated temperatures and a strong increase for $T<50$\,K. We fit the $\tau$ values for $T\leq50$\,K to Eq.\,(\ref{Eq: time}) and extract $E_{\mathrm{A}}=(6.2\pm0.3)$\,meV and $\tau_0=(19.3\pm0.2)$\,ps. The value of $\tau_0$ is in good agreement to the fitting results from Fig.\,\ref{Fig: 5}(d) from the main text, while the fitted activation energy $E_{\mathrm{A}}$ is larger than for the thicker maghemite sample.
\bibliography{library}

\begin{thebibliography}{60}%
\makeatletter
\providecommand \@ifxundefined [1]{%
 \@ifx{#1\undefined}
}%
\providecommand \@ifnum [1]{%
 \ifnum #1\expandafter \@firstoftwo
 \else \expandafter \@secondoftwo
 \fi
}%
\providecommand \@ifx [1]{%
 \ifx #1\expandafter \@firstoftwo
 \else \expandafter \@secondoftwo
 \fi
}%
\providecommand \natexlab [1]{#1}%
\providecommand \enquote  [1]{``#1''}%
\providecommand \bibnamefont  [1]{#1}%
\providecommand \bibfnamefont [1]{#1}%
\providecommand \citenamefont [1]{#1}%
\providecommand \href@noop [0]{\@secondoftwo}%
\providecommand \href [0]{\begingroup \@sanitize@url \@href}%
\providecommand \@href[1]{\@@startlink{#1}\@@href}%
\providecommand \@@href[1]{\endgroup#1\@@endlink}%
\providecommand \@sanitize@url [0]{\catcode `\\12\catcode `\$12\catcode
  `\&12\catcode `\#12\catcode `\^12\catcode `\_12\catcode `\%12\relax}%
\providecommand \@@startlink[1]{}%
\providecommand \@@endlink[0]{}%
\providecommand \url  [0]{\begingroup\@sanitize@url \@url }%
\providecommand \@url [1]{\endgroup\@href {#1}{\urlprefix }}%
\providecommand \urlprefix  [0]{URL }%
\providecommand \Eprint [0]{\href }%
\providecommand \doibase [0]{https://doi.org/}%
\providecommand \selectlanguage [0]{\@gobble}%
\providecommand \bibinfo  [0]{\@secondoftwo}%
\providecommand \bibfield  [0]{\@secondoftwo}%
\providecommand \translation [1]{[#1]}%
\providecommand \BibitemOpen [0]{}%
\providecommand \bibitemStop [0]{}%
\providecommand \bibitemNoStop [0]{.\EOS\space}%
\providecommand \EOS [0]{\spacefactor3000\relax}%
\providecommand \BibitemShut  [1]{\csname bibitem#1\endcsname}%
\let\auto@bib@innerbib\@empty
\bibitem [{\citenamefont {Brataas}\ \emph {et~al.}(2020)\citenamefont
  {Brataas}, \citenamefont {van Wees}, \citenamefont {Klein}, \citenamefont
  {de~Loubens},\ and\ \citenamefont {Viret}}]{Brataas2020a}%
  \BibitemOpen
  \bibfield  {author} {\bibinfo {author} {\bibfnamefont {A.}~\bibnamefont
  {Brataas}}, \bibinfo {author} {\bibfnamefont {B.}~\bibnamefont {van Wees}},
  \bibinfo {author} {\bibfnamefont {O.}~\bibnamefont {Klein}}, \bibinfo
  {author} {\bibfnamefont {G.}~\bibnamefont {de~Loubens}},\ and\ \bibinfo
  {author} {\bibfnamefont {M.}~\bibnamefont {Viret}},\ }\bibfield  {title}
  {\bibinfo {title} {{Spin insulatronics}},\ }\href
  {https://doi.org/10.1016/j.physrep.2020.08.006} {\bibfield  {journal}
  {\bibinfo  {journal} {Phys. Rep.}\ }\textbf {\bibinfo {volume} {885}},\
  \bibinfo {pages} {1} (\bibinfo {year} {2020})}\BibitemShut {NoStop}%
\bibitem [{\citenamefont {Chumak}\ \emph {et~al.}(2015)\citenamefont {Chumak},
  \citenamefont {Vasyuchka}, \citenamefont {Serga},\ and\ \citenamefont
  {Hillebrands}}]{Chumak2015}%
  \BibitemOpen
  \bibfield  {author} {\bibinfo {author} {\bibfnamefont {A.~V.}\ \bibnamefont
  {Chumak}}, \bibinfo {author} {\bibfnamefont {V.~I.}\ \bibnamefont
  {Vasyuchka}}, \bibinfo {author} {\bibfnamefont {A.~A.}\ \bibnamefont
  {Serga}},\ and\ \bibinfo {author} {\bibfnamefont {B.}~\bibnamefont
  {Hillebrands}},\ }\bibfield  {title} {\bibinfo {title} {{Magnon
  spintronics}},\ }\href {https://doi.org/10.1038/nphys3347} {\bibfield
  {journal} {\bibinfo  {journal} {Nature Physics}\ }\textbf {\bibinfo {volume}
  {11}},\ \bibinfo {pages} {453} (\bibinfo {year} {2015})}\BibitemShut
  {NoStop}%
\bibitem [{\citenamefont {Chumak}\ \emph {et~al.}(2017)\citenamefont {Chumak},
  \citenamefont {Serga},\ and\ \citenamefont {Hillebrands}}]{Chumak2017}%
  \BibitemOpen
  \bibfield  {author} {\bibinfo {author} {\bibfnamefont {A.~V.}\ \bibnamefont
  {Chumak}}, \bibinfo {author} {\bibfnamefont {A.~A.}\ \bibnamefont {Serga}},\
  and\ \bibinfo {author} {\bibfnamefont {B.}~\bibnamefont {Hillebrands}},\
  }\bibfield  {title} {\bibinfo {title} {{Magnonic crystals for data
  processing}},\ }\href {https://doi.org/10.1088/1361-6463/aa6a65} {\bibfield
  {journal} {\bibinfo  {journal} {J. Phys. D: Appl. Phys}\ }\textbf {\bibinfo
  {volume} {50}},\ \bibinfo {pages} {244001} (\bibinfo {year}
  {2017})}\BibitemShut {NoStop}%
\bibitem [{\citenamefont {Bauer}\ \emph {et~al.}(2012)\citenamefont {Bauer},
  \citenamefont {Saitoh},\ and\ \citenamefont {{Van Wees}}}]{Bauer2012}%
  \BibitemOpen
  \bibfield  {author} {\bibinfo {author} {\bibfnamefont {G.~E.}\ \bibnamefont
  {Bauer}}, \bibinfo {author} {\bibfnamefont {E.}~\bibnamefont {Saitoh}},\ and\
  \bibinfo {author} {\bibfnamefont {B.~J.}\ \bibnamefont {{Van Wees}}},\
  }\bibfield  {title} {\bibinfo {title} {{Spin caloritronics}},\ }\href
  {https://doi.org/10.1038/nmat3301} {\bibfield  {journal} {\bibinfo  {journal}
  {Nat. Mater.}\ }\textbf {\bibinfo {volume} {11}},\ \bibinfo {pages} {391}
  (\bibinfo {year} {2012})}\BibitemShut {NoStop}%
\bibitem [{\citenamefont {Li}\ \emph {et~al.}(2016)\citenamefont {Li},
  \citenamefont {Liu}, \citenamefont {Chang}, \citenamefont {Kalitsov},
  \citenamefont {Zhang}, \citenamefont {Csaba}, \citenamefont {Li},
  \citenamefont {Richardson}, \citenamefont {DeMann}, \citenamefont {Rimal},
  \citenamefont {Dey}, \citenamefont {Jiang}, \citenamefont {Porod},
  \citenamefont {Field}, \citenamefont {Tang}, \citenamefont {Marconi},
  \citenamefont {Hoffmann}, \citenamefont {Mryasov},\ and\ \citenamefont
  {Wu}}]{Li2016}%
  \BibitemOpen
  \bibfield  {author} {\bibinfo {author} {\bibfnamefont {P.}~\bibnamefont
  {Li}}, \bibinfo {author} {\bibfnamefont {T.}~\bibnamefont {Liu}}, \bibinfo
  {author} {\bibfnamefont {H.}~\bibnamefont {Chang}}, \bibinfo {author}
  {\bibfnamefont {A.}~\bibnamefont {Kalitsov}}, \bibinfo {author}
  {\bibfnamefont {W.}~\bibnamefont {Zhang}}, \bibinfo {author} {\bibfnamefont
  {G.}~\bibnamefont {Csaba}}, \bibinfo {author} {\bibfnamefont
  {W.}~\bibnamefont {Li}}, \bibinfo {author} {\bibfnamefont {D.}~\bibnamefont
  {Richardson}}, \bibinfo {author} {\bibfnamefont {A.}~\bibnamefont {DeMann}},
  \bibinfo {author} {\bibfnamefont {G.}~\bibnamefont {Rimal}}, \bibinfo
  {author} {\bibfnamefont {H.}~\bibnamefont {Dey}}, \bibinfo {author}
  {\bibfnamefont {J.~S.}\ \bibnamefont {Jiang}}, \bibinfo {author}
  {\bibfnamefont {W.}~\bibnamefont {Porod}}, \bibinfo {author} {\bibfnamefont
  {S.~B.}\ \bibnamefont {Field}}, \bibinfo {author} {\bibfnamefont
  {J.}~\bibnamefont {Tang}}, \bibinfo {author} {\bibfnamefont {M.~C.}\
  \bibnamefont {Marconi}}, \bibinfo {author} {\bibfnamefont {A.}~\bibnamefont
  {Hoffmann}}, \bibinfo {author} {\bibfnamefont {O.}~\bibnamefont {Mryasov}},\
  and\ \bibinfo {author} {\bibfnamefont {M.}~\bibnamefont {Wu}},\ }\bibfield
  {title} {\bibinfo {title} {{Spin-orbit torque-assisted switching in magnetic
  insulator thin films with perpendicular magnetic anisotropy}},\ }\href
  {https://doi.org/10.1038/ncomms12688} {\bibfield  {journal} {\bibinfo
  {journal} {Nat. Commun.}\ }\textbf {\bibinfo {volume} {7}},\ \bibinfo {pages}
  {1} (\bibinfo {year} {2016})}\BibitemShut {NoStop}%
\bibitem [{\citenamefont {Cherepanov}\ \emph {et~al.}(1993)\citenamefont
  {Cherepanov}, \citenamefont {Kolokolov},\ and\ \citenamefont
  {L'vov}}]{Cherepanov1993}%
  \BibitemOpen
  \bibfield  {author} {\bibinfo {author} {\bibfnamefont {V.}~\bibnamefont
  {Cherepanov}}, \bibinfo {author} {\bibfnamefont {I.}~\bibnamefont
  {Kolokolov}},\ and\ \bibinfo {author} {\bibfnamefont {V.}~\bibnamefont
  {L'vov}},\ }\bibfield  {title} {\bibinfo {title} {{The saga of YIG: Spectra,
  thermodynamics, interaction and relaxation of magnons in a complex magnet}},\
  }\href {https://doi.org/10.1016/0370-1573(93)90107-O} {\bibfield  {journal}
  {\bibinfo  {journal} {Phys. Rep.}\ }\textbf {\bibinfo {volume} {229}},\
  \bibinfo {pages} {81} (\bibinfo {year} {1993})}\BibitemShut {NoStop}%
\bibitem [{\citenamefont {Spencer}\ \emph {et~al.}(1959)\citenamefont
  {Spencer}, \citenamefont {LeCraw},\ and\ \citenamefont
  {Clogston}}]{Spencer1959}%
  \BibitemOpen
  \bibfield  {author} {\bibinfo {author} {\bibfnamefont {E.~G.}\ \bibnamefont
  {Spencer}}, \bibinfo {author} {\bibfnamefont {R.~C.}\ \bibnamefont
  {LeCraw}},\ and\ \bibinfo {author} {\bibfnamefont {A.~M.}\ \bibnamefont
  {Clogston}},\ }\bibfield  {title} {\bibinfo {title} {{Low-Temperature
  Line-Width Maximum in Yttrium Iron Garnet}},\ }\href
  {https://doi.org/10.1103/PhysRevLett.3.32} {\bibfield  {journal} {\bibinfo
  {journal} {Phys. Rev. Lett.}\ }\textbf {\bibinfo {volume} {3}},\ \bibinfo
  {pages} {32} (\bibinfo {year} {1959})}\BibitemShut {NoStop}%
\bibitem [{\citenamefont {Maier-Flaig}\ \emph {et~al.}(2017)\citenamefont
  {Maier-Flaig}, \citenamefont {Klingler}, \citenamefont {Dubs}, \citenamefont
  {Surzhenko}, \citenamefont {Gross}, \citenamefont {Weiler}, \citenamefont
  {Huebl},\ and\ \citenamefont {Goennenwein}}]{Maier-Flaig2017}%
  \BibitemOpen
  \bibfield  {author} {\bibinfo {author} {\bibfnamefont {H.}~\bibnamefont
  {Maier-Flaig}}, \bibinfo {author} {\bibfnamefont {S.}~\bibnamefont
  {Klingler}}, \bibinfo {author} {\bibfnamefont {C.}~\bibnamefont {Dubs}},
  \bibinfo {author} {\bibfnamefont {O.}~\bibnamefont {Surzhenko}}, \bibinfo
  {author} {\bibfnamefont {R.}~\bibnamefont {Gross}}, \bibinfo {author}
  {\bibfnamefont {M.}~\bibnamefont {Weiler}}, \bibinfo {author} {\bibfnamefont
  {H.}~\bibnamefont {Huebl}},\ and\ \bibinfo {author} {\bibfnamefont {S.~T.}\
  \bibnamefont {Goennenwein}},\ }\bibfield  {title} {\bibinfo {title}
  {{Temperature-dependent magnetic damping of yttrium iron garnet spheres}},\
  }\href {https://doi.org/10.1103/PhysRevB.95.214423} {\bibfield  {journal}
  {\bibinfo  {journal} {Phys. Rev. B}\ }\textbf {\bibinfo {volume} {95}},\
  \bibinfo {pages} {1} (\bibinfo {year} {2017})}\BibitemShut {NoStop}%
\bibitem [{\citenamefont {Klingler}\ \emph {et~al.}(2017)\citenamefont
  {Klingler}, \citenamefont {Maier-Flaig}, \citenamefont {Dubs}, \citenamefont
  {Surzhenko}, \citenamefont {Gross}, \citenamefont {Huebl}, \citenamefont
  {Goennenwein},\ and\ \citenamefont {Weiler}}]{Klingler2017}%
  \BibitemOpen
  \bibfield  {author} {\bibinfo {author} {\bibfnamefont {S.}~\bibnamefont
  {Klingler}}, \bibinfo {author} {\bibfnamefont {H.}~\bibnamefont
  {Maier-Flaig}}, \bibinfo {author} {\bibfnamefont {C.}~\bibnamefont {Dubs}},
  \bibinfo {author} {\bibfnamefont {O.}~\bibnamefont {Surzhenko}}, \bibinfo
  {author} {\bibfnamefont {R.}~\bibnamefont {Gross}}, \bibinfo {author}
  {\bibfnamefont {H.}~\bibnamefont {Huebl}}, \bibinfo {author} {\bibfnamefont
  {S.~T.~B.}\ \bibnamefont {Goennenwein}},\ and\ \bibinfo {author}
  {\bibfnamefont {M.}~\bibnamefont {Weiler}},\ }\bibfield  {title}
  {{\selectlanguage {English}\bibinfo {title} {{Gilbert damping of
  magnetostatic modes in a yttrium iron garnet sphere}}},\ }\href
  {https://doi.org/10.1063/1.4977423} {\bibfield  {journal} {\bibinfo
  {journal} {Applied Physics Letters}\ }\textbf {\bibinfo {volume} {110}},\
  \bibinfo {pages} {092409} (\bibinfo {year} {2017})}\BibitemShut {NoStop}%
\bibitem [{\citenamefont {Serga}\ \emph {et~al.}(2010)\citenamefont {Serga},
  \citenamefont {Chumak},\ and\ \citenamefont {Hillebrands}}]{Serga2010}%
  \BibitemOpen
  \bibfield  {author} {\bibinfo {author} {\bibfnamefont {A.~A.}\ \bibnamefont
  {Serga}}, \bibinfo {author} {\bibfnamefont {A.~V.}\ \bibnamefont {Chumak}},\
  and\ \bibinfo {author} {\bibfnamefont {B.}~\bibnamefont {Hillebrands}},\
  }\bibfield  {title} {\bibinfo {title} {{YIG magnonics}},\ }\href
  {https://doi.org/10.1088/0022-3727/43/26/264002} {\bibfield  {journal}
  {\bibinfo  {journal} {J. Phys. D: Appl. Phys}\ }\textbf {\bibinfo {volume}
  {43}},\ \bibinfo {pages} {264002} (\bibinfo {year} {2010})}\BibitemShut
  {NoStop}%
\bibitem [{\citenamefont {Seifert}\ \emph {et~al.}(2018)\citenamefont
  {Seifert}, \citenamefont {Jaiswal}, \citenamefont {Barker}, \citenamefont
  {Weber}, \citenamefont {Razdolski}, \citenamefont {Cramer}, \citenamefont
  {Gueckstock}, \citenamefont {Maehrlein}, \citenamefont {Nadvornik},
  \citenamefont {Watanabe}, \citenamefont {Ciccarelli}, \citenamefont
  {Melnikov}, \citenamefont {Jakob}, \citenamefont {M{\"{u}}nzenberg},
  \citenamefont {Goennenwein}, \citenamefont {Woltersdorf}, \citenamefont
  {Rethfeld}, \citenamefont {Brouwer}, \citenamefont {Wolf}, \citenamefont
  {Kl{\"{a}}ui},\ and\ \citenamefont {Kampfrath}}]{Seifert2018}%
  \BibitemOpen
  \bibfield  {author} {\bibinfo {author} {\bibfnamefont {T.~S.}\ \bibnamefont
  {Seifert}}, \bibinfo {author} {\bibfnamefont {S.}~\bibnamefont {Jaiswal}},
  \bibinfo {author} {\bibfnamefont {J.}~\bibnamefont {Barker}}, \bibinfo
  {author} {\bibfnamefont {S.~T.}\ \bibnamefont {Weber}}, \bibinfo {author}
  {\bibfnamefont {I.}~\bibnamefont {Razdolski}}, \bibinfo {author}
  {\bibfnamefont {J.}~\bibnamefont {Cramer}}, \bibinfo {author} {\bibfnamefont
  {O.}~\bibnamefont {Gueckstock}}, \bibinfo {author} {\bibfnamefont {S.~F.}\
  \bibnamefont {Maehrlein}}, \bibinfo {author} {\bibfnamefont {L.}~\bibnamefont
  {Nadvornik}}, \bibinfo {author} {\bibfnamefont {S.}~\bibnamefont {Watanabe}},
  \bibinfo {author} {\bibfnamefont {C.}~\bibnamefont {Ciccarelli}}, \bibinfo
  {author} {\bibfnamefont {A.}~\bibnamefont {Melnikov}}, \bibinfo {author}
  {\bibfnamefont {G.}~\bibnamefont {Jakob}}, \bibinfo {author} {\bibfnamefont
  {M.}~\bibnamefont {M{\"{u}}nzenberg}}, \bibinfo {author} {\bibfnamefont
  {S.~T.~B.}\ \bibnamefont {Goennenwein}}, \bibinfo {author} {\bibfnamefont
  {G.}~\bibnamefont {Woltersdorf}}, \bibinfo {author} {\bibfnamefont
  {B.}~\bibnamefont {Rethfeld}}, \bibinfo {author} {\bibfnamefont {P.~W.}\
  \bibnamefont {Brouwer}}, \bibinfo {author} {\bibfnamefont {M.}~\bibnamefont
  {Wolf}}, \bibinfo {author} {\bibfnamefont {M.}~\bibnamefont {Kl{\"{a}}ui}},\
  and\ \bibinfo {author} {\bibfnamefont {T.}~\bibnamefont {Kampfrath}},\
  }\bibfield  {title} {\bibinfo {title} {{Femtosecond formation dynamics of the
  spin Seebeck effect revealed by terahertz spectroscopy}},\ }\href
  {https://doi.org/10.1038/s41467-018-05135-2} {\bibfield  {journal} {\bibinfo
  {journal} {Nat. Commun.}\ }\textbf {\bibinfo {volume} {9}},\ \bibinfo {pages}
  {2899} (\bibinfo {year} {2018})}\BibitemShut {NoStop}%
\bibitem [{\citenamefont {Grau-Crespo}\ \emph {et~al.}(2010)\citenamefont
  {Grau-Crespo}, \citenamefont {Al-Baitai}, \citenamefont {Saadoune},\ and\
  \citenamefont {{De Leeuw}}}]{Grau-Crespo2010}%
  \BibitemOpen
  \bibfield  {author} {\bibinfo {author} {\bibfnamefont {R.}~\bibnamefont
  {Grau-Crespo}}, \bibinfo {author} {\bibfnamefont {A.~Y.}\ \bibnamefont
  {Al-Baitai}}, \bibinfo {author} {\bibfnamefont {I.}~\bibnamefont
  {Saadoune}},\ and\ \bibinfo {author} {\bibfnamefont {N.~H.}\ \bibnamefont
  {{De Leeuw}}},\ }\bibfield  {title} {\bibinfo {title} {{Vacancy ordering and
  electronic structure of $\gamma$ -\ch{Fe2O3} (maghemite): a theoretical
  investigation}},\ }\href {https://doi.org/10.1088/0953-8984/22/25/255401}
  {\bibfield  {journal} {\bibinfo  {journal} {J. Phys. Condens. Matter}\
  }\textbf {\bibinfo {volume} {22}},\ \bibinfo {pages} {255401} (\bibinfo
  {year} {2010})}\BibitemShut {NoStop}%
\bibitem [{\citenamefont {Huang}\ \emph {et~al.}(2013)\citenamefont {Huang},
  \citenamefont {Yang},\ and\ \citenamefont {Ding}}]{Huang2013}%
  \BibitemOpen
  \bibfield  {author} {\bibinfo {author} {\bibfnamefont {X.}~\bibnamefont
  {Huang}}, \bibinfo {author} {\bibfnamefont {Y.}~\bibnamefont {Yang}},\ and\
  \bibinfo {author} {\bibfnamefont {J.}~\bibnamefont {Ding}},\ }\bibfield
  {title} {\bibinfo {title} {{Epitaxial growth of $\gamma$-\ch{Fe2O3} thin
  films on MgO substrates by pulsed laser deposition and their properties}},\
  }\href {https://doi.org/10.1016/j.actamat.2012.10.003} {\bibfield  {journal}
  {\bibinfo  {journal} {Acta Mater.}\ }\textbf {\bibinfo {volume} {61}},\
  \bibinfo {pages} {548} (\bibinfo {year} {2013})}\BibitemShut {NoStop}%
\bibitem [{\citenamefont {Dronskowski}(2001)}]{Dronskowski2001}%
  \BibitemOpen
  \bibfield  {author} {\bibinfo {author} {\bibfnamefont {R.}~\bibnamefont
  {Dronskowski}},\ }\bibfield  {title} {\bibinfo {title} {{The Little Maghemite
  Story: A Classic Functional Material}},\ }\href
  {https://doi.org/10.1002/1616-3028(200102)11:1<27::AID-ADFM27>3.0.CO;2-X}
  {\bibfield  {journal} {\bibinfo  {journal} {Adv. Funct. Mater.}\ }\textbf
  {\bibinfo {volume} {11}},\ \bibinfo {pages} {27} (\bibinfo {year}
  {2001})}\BibitemShut {NoStop}%
\bibitem [{\citenamefont {P{\'{a}}lfalvi}\ \emph {et~al.}(2005)\citenamefont
  {P{\'{a}}lfalvi}, \citenamefont {Hebling}, \citenamefont {Kuhl},
  \citenamefont {P{\'{e}}ter},\ and\ \citenamefont
  {Polg{\'{a}}r}}]{Palfalvi2005}%
  \BibitemOpen
  \bibfield  {author} {\bibinfo {author} {\bibfnamefont {L.}~\bibnamefont
  {P{\'{a}}lfalvi}}, \bibinfo {author} {\bibfnamefont {J.}~\bibnamefont
  {Hebling}}, \bibinfo {author} {\bibfnamefont {J.}~\bibnamefont {Kuhl}},
  \bibinfo {author} {\bibfnamefont {{\'{A}}.}~\bibnamefont {P{\'{e}}ter}},\
  and\ \bibinfo {author} {\bibfnamefont {K.}~\bibnamefont {Polg{\'{a}}r}},\
  }\bibfield  {title} {\bibinfo {title} {{Temperature dependence of the
  absorption and refraction of Mg-doped congruent and stoichiometric
  $\mathrm{LiNbO_3}$ in the THz range}},\ }\href
  {https://doi.org/10.1063/1.1929859} {\bibfield  {journal} {\bibinfo
  {journal} {Journ. Appl. Phys.}\ }\textbf {\bibinfo {volume} {97}},\ \bibinfo
  {pages} {123505} (\bibinfo {year} {2005})}\BibitemShut {NoStop}%
\bibitem [{\citenamefont {Jim{\'{e}}nez-Cavero}\ \emph
  {et~al.}(2017)\citenamefont {Jim{\'{e}}nez-Cavero}, \citenamefont {Lucas},
  \citenamefont {Anad{\'{o}}n}, \citenamefont {Ramos}, \citenamefont {Niizeki},
  \citenamefont {Aguirre}, \citenamefont {Algarabel}, \citenamefont {Uchida},
  \citenamefont {Ibarra}, \citenamefont {Saitoh},\ and\ \citenamefont
  {Morell{\'{o}}n}}]{Jimenez-Cavero2017}%
  \BibitemOpen
  \bibfield  {author} {\bibinfo {author} {\bibfnamefont {P.}~\bibnamefont
  {Jim{\'{e}}nez-Cavero}}, \bibinfo {author} {\bibfnamefont {I.}~\bibnamefont
  {Lucas}}, \bibinfo {author} {\bibfnamefont {A.}~\bibnamefont {Anad{\'{o}}n}},
  \bibinfo {author} {\bibfnamefont {R.}~\bibnamefont {Ramos}}, \bibinfo
  {author} {\bibfnamefont {T.}~\bibnamefont {Niizeki}}, \bibinfo {author}
  {\bibfnamefont {M.~H.}\ \bibnamefont {Aguirre}}, \bibinfo {author}
  {\bibfnamefont {P.~A.}\ \bibnamefont {Algarabel}}, \bibinfo {author}
  {\bibfnamefont {K.}~\bibnamefont {Uchida}}, \bibinfo {author} {\bibfnamefont
  {M.~R.}\ \bibnamefont {Ibarra}}, \bibinfo {author} {\bibfnamefont
  {E.}~\bibnamefont {Saitoh}},\ and\ \bibinfo {author} {\bibfnamefont
  {L.}~\bibnamefont {Morell{\'{o}}n}},\ }\bibfield  {title} {\bibinfo {title}
  {{Spin Seebeck effect in insulating epitaxial $\gamma$-$\mathrm{Fe_2O_3}$
  thin films}},\ }\href {http://dx.doi.org/10.1063/1.4975618} {\bibfield
  {journal} {\bibinfo  {journal} {APL Mater.}\ }\textbf {\bibinfo {volume} {5}}
  (\bibinfo {year} {2017})}\BibitemShut {NoStop}%
\bibitem [{\citenamefont {Lee}\ \emph {et~al.}(2001)\citenamefont {Lee},
  \citenamefont {Jang}, \citenamefont {Kim},\ and\ \citenamefont
  {Yoon}}]{Lee2001}%
  \BibitemOpen
  \bibfield  {author} {\bibinfo {author} {\bibfnamefont {E.-T.}\ \bibnamefont
  {Lee}}, \bibinfo {author} {\bibfnamefont {G.-E.}\ \bibnamefont {Jang}},
  \bibinfo {author} {\bibfnamefont {C.~K.}\ \bibnamefont {Kim}},\ and\ \bibinfo
  {author} {\bibfnamefont {D.-H.}\ \bibnamefont {Yoon}},\ }\bibfield  {title}
  {\bibinfo {title} {{Fabrication and gas sensing properties of
  $\alpha$-$\mathrm{Fe_2O_3}$ thin film prepared by plasma enhanced chemical
  vapor deposition (PECVD)}},\ }\href
  {https://doi.org/10.1016/S0925-4005(01)00716-X} {\bibfield  {journal}
  {\bibinfo  {journal} {Sens. Actuators B Chem.}\ }\textbf {\bibinfo {volume}
  {77}},\ \bibinfo {pages} {221} (\bibinfo {year} {2001})}\BibitemShut
  {NoStop}%
\bibitem [{\citenamefont {Dghoughi}\ \emph {et~al.}(2006)\citenamefont
  {Dghoughi}, \citenamefont {Elidrissi}, \citenamefont {Bern{\`{e}}de},
  \citenamefont {Addou}, \citenamefont {Lamrani}, \citenamefont {Regragui},\
  and\ \citenamefont {Erguig}}]{Dghoughi2006}%
  \BibitemOpen
  \bibfield  {author} {\bibinfo {author} {\bibfnamefont {L.}~\bibnamefont
  {Dghoughi}}, \bibinfo {author} {\bibfnamefont {B.}~\bibnamefont {Elidrissi}},
  \bibinfo {author} {\bibfnamefont {C.}~\bibnamefont {Bern{\`{e}}de}}, \bibinfo
  {author} {\bibfnamefont {M.}~\bibnamefont {Addou}}, \bibinfo {author}
  {\bibfnamefont {M.~A.}\ \bibnamefont {Lamrani}}, \bibinfo {author}
  {\bibfnamefont {M.}~\bibnamefont {Regragui}},\ and\ \bibinfo {author}
  {\bibfnamefont {H.}~\bibnamefont {Erguig}},\ }\bibfield  {title} {\bibinfo
  {title} {{Physico-chemical, optical and electrochemical properties of iron
  oxide thin films prepared by spray pyrolysis}},\ }\href
  {https://doi.org/10.1016/j.apsusc.2006.03.021} {\bibfield  {journal}
  {\bibinfo  {journal} {Appl. Surf. Sci.}\ }\textbf {\bibinfo {volume} {253}},\
  \bibinfo {pages} {1823} (\bibinfo {year} {2006})}\BibitemShut {NoStop}%
\bibitem [{\citenamefont {Alraddadi}(2020)}]{Alraddadi2020}%
  \BibitemOpen
  \bibfield  {author} {\bibinfo {author} {\bibfnamefont {S.}~\bibnamefont
  {Alraddadi}},\ }\bibfield  {title} {\bibinfo {title} {{The Electronic and
  Magnetic Properties of Ultrathin $\gamma$-\ch{Fe2O3} Films}},\ }\href
  {https://doi.org/10.1088/1757-899X/842/1/012012} {\bibfield  {journal}
  {\bibinfo  {journal} {IOP Conf. Ser.: Mater. Sci. Eng.}\ }\textbf {\bibinfo
  {volume} {842}},\ \bibinfo {pages} {012012} (\bibinfo {year}
  {2020})}\BibitemShut {NoStop}%
\bibitem [{\citenamefont {G{\"{u}}ckelhorn}\ \emph {et~al.}(2021)\citenamefont
  {G{\"{u}}ckelhorn}, \citenamefont {Wimmer}, \citenamefont {M{\"{u}}ller},
  \citenamefont {Gepr{\"{a}}gs}, \citenamefont {Huebl}, \citenamefont {Gross},\
  and\ \citenamefont {Althammer}}]{Guckelhorn2021}%
  \BibitemOpen
  \bibfield  {author} {\bibinfo {author} {\bibfnamefont {J.}~\bibnamefont
  {G{\"{u}}ckelhorn}}, \bibinfo {author} {\bibfnamefont {T.}~\bibnamefont
  {Wimmer}}, \bibinfo {author} {\bibfnamefont {M.}~\bibnamefont
  {M{\"{u}}ller}}, \bibinfo {author} {\bibfnamefont {S.}~\bibnamefont
  {Gepr{\"{a}}gs}}, \bibinfo {author} {\bibfnamefont {H.}~\bibnamefont
  {Huebl}}, \bibinfo {author} {\bibfnamefont {R.}~\bibnamefont {Gross}},\ and\
  \bibinfo {author} {\bibfnamefont {M.}~\bibnamefont {Althammer}},\ }\bibfield
  {title} {\bibinfo {title} {{Magnon transport in {\ch{Y3Fe5O12}/Pt
  nanostructures with reduced effective magnetization}}},\ }\href
  {https://doi.org/10.1103/PhysRevB.104.L180410} {\bibfield  {journal}
  {\bibinfo  {journal} {Phys. Rev. B}\ }\textbf {\bibinfo {volume} {104}},\
  \bibinfo {pages} {L180410} (\bibinfo {year} {2021})}\BibitemShut {NoStop}%
\bibitem [{\citenamefont {Divinskiy}\ \emph {et~al.}(2021)\citenamefont
  {Divinskiy}, \citenamefont {Merbouche}, \citenamefont {Demidov},
  \citenamefont {Nikolaev}, \citenamefont {Soumah}, \citenamefont
  {Gou{\'{e}}r{\'{e}}}, \citenamefont {Lebrun}, \citenamefont {Cros},
  \citenamefont {Youssef}, \citenamefont {Bortolotti}, \citenamefont {Anane},\
  and\ \citenamefont {Demokritov}}]{Divinskiy2021}%
  \BibitemOpen
  \bibfield  {author} {\bibinfo {author} {\bibfnamefont {B.}~\bibnamefont
  {Divinskiy}}, \bibinfo {author} {\bibfnamefont {H.}~\bibnamefont
  {Merbouche}}, \bibinfo {author} {\bibfnamefont {V.~E.}\ \bibnamefont
  {Demidov}}, \bibinfo {author} {\bibfnamefont {K.~O.}\ \bibnamefont
  {Nikolaev}}, \bibinfo {author} {\bibfnamefont {L.}~\bibnamefont {Soumah}},
  \bibinfo {author} {\bibfnamefont {D.}~\bibnamefont {Gou{\'{e}}r{\'{e}}}},
  \bibinfo {author} {\bibfnamefont {R.}~\bibnamefont {Lebrun}}, \bibinfo
  {author} {\bibfnamefont {V.}~\bibnamefont {Cros}}, \bibinfo {author}
  {\bibfnamefont {J.~B.}\ \bibnamefont {Youssef}}, \bibinfo {author}
  {\bibfnamefont {P.}~\bibnamefont {Bortolotti}}, \bibinfo {author}
  {\bibfnamefont {A.}~\bibnamefont {Anane}},\ and\ \bibinfo {author}
  {\bibfnamefont {S.~O.}\ \bibnamefont {Demokritov}},\ }\bibfield  {title}
  {\bibinfo {title} {{Evidence for spin current driven Bose-Einstein
  condensation of magnons}},\ }\href
  {https://doi.org/10.1038/s41467-021-26790-y} {\bibfield  {journal} {\bibinfo
  {journal} {Nat. Commun.}\ }\textbf {\bibinfo {volume} {12}},\ \bibinfo
  {pages} {6541} (\bibinfo {year} {2021})}\BibitemShut {NoStop}%
\bibitem [{\citenamefont {Evelt}\ \emph {et~al.}(2018)\citenamefont {Evelt},
  \citenamefont {Soumah}, \citenamefont {Rinkevich}, \citenamefont
  {Demokritov}, \citenamefont {Anane}, \citenamefont {Cros}, \citenamefont
  {{Ben Youssef}}, \citenamefont {de~Loubens}, \citenamefont {Klein},
  \citenamefont {Bortolotti},\ and\ \citenamefont {Demidov}}]{Evelt2018}%
  \BibitemOpen
  \bibfield  {author} {\bibinfo {author} {\bibfnamefont {M.}~\bibnamefont
  {Evelt}}, \bibinfo {author} {\bibfnamefont {L.}~\bibnamefont {Soumah}},
  \bibinfo {author} {\bibfnamefont {A.}~\bibnamefont {Rinkevich}}, \bibinfo
  {author} {\bibfnamefont {S.}~\bibnamefont {Demokritov}}, \bibinfo {author}
  {\bibfnamefont {A.}~\bibnamefont {Anane}}, \bibinfo {author} {\bibfnamefont
  {V.}~\bibnamefont {Cros}}, \bibinfo {author} {\bibfnamefont {J.}~\bibnamefont
  {{Ben Youssef}}}, \bibinfo {author} {\bibfnamefont {G.}~\bibnamefont
  {de~Loubens}}, \bibinfo {author} {\bibfnamefont {O.}~\bibnamefont {Klein}},
  \bibinfo {author} {\bibfnamefont {P.}~\bibnamefont {Bortolotti}},\ and\
  \bibinfo {author} {\bibfnamefont {V.}~\bibnamefont {Demidov}},\ }\bibfield
  {title} {\bibinfo {title} {{Emission of Coherent Propagating Magnons by
  Insulator-Based Spin-Orbit-Torque Oscillators}},\ }\href
  {https://doi.org/10.1103/PhysRevApplied.10.041002} {\bibfield  {journal}
  {\bibinfo  {journal} {Phys. Rev. Appl.}\ }\textbf {\bibinfo {volume} {10}},\
  \bibinfo {pages} {041002} (\bibinfo {year} {2018})}\BibitemShut {NoStop}%
\bibitem [{\citenamefont {{Van Vleck}}\ and\ \citenamefont
  {Orbach}(1963)}]{VanVleck1963}%
  \BibitemOpen
  \bibfield  {author} {\bibinfo {author} {\bibfnamefont {J.~H.}\ \bibnamefont
  {{Van Vleck}}}\ and\ \bibinfo {author} {\bibfnamefont {R.}~\bibnamefont
  {Orbach}},\ }\bibfield  {title} {\bibinfo {title} {{Ferrimagnetic Resonance
  of Dilute Rare-Earth Doped Iron Garnets}},\ }\href
  {https://doi.org/10.1103/PhysRevLett.11.65} {\bibfield  {journal} {\bibinfo
  {journal} {Phys. Rev. Lett.}\ }\textbf {\bibinfo {volume} {11}},\ \bibinfo
  {pages} {65} (\bibinfo {year} {1963})}\BibitemShut {NoStop}%
\bibitem [{\citenamefont {Nembach}\ \emph {et~al.}(2011)\citenamefont
  {Nembach}, \citenamefont {Silva}, \citenamefont {Shaw}, \citenamefont
  {Schneider}, \citenamefont {Carey}, \citenamefont {Maat},\ and\ \citenamefont
  {Childress}}]{Nembach2011}%
  \BibitemOpen
  \bibfield  {author} {\bibinfo {author} {\bibfnamefont {H.~T.}\ \bibnamefont
  {Nembach}}, \bibinfo {author} {\bibfnamefont {T.~J.}\ \bibnamefont {Silva}},
  \bibinfo {author} {\bibfnamefont {J.~M.}\ \bibnamefont {Shaw}}, \bibinfo
  {author} {\bibfnamefont {M.~L.}\ \bibnamefont {Schneider}}, \bibinfo {author}
  {\bibfnamefont {M.~J.}\ \bibnamefont {Carey}}, \bibinfo {author}
  {\bibfnamefont {S.}~\bibnamefont {Maat}},\ and\ \bibinfo {author}
  {\bibfnamefont {J.~R.}\ \bibnamefont {Childress}},\ }\bibfield  {title}
  {\bibinfo {title} {{Perpendicular ferromagnetic resonance measurements of
  damping and Lande g-factor in sputtered $\mathrm{(Co_2Mn)_{1-x}Ge_x}$ thin
  films}},\ }\href {https://doi.org/10.1103/PhysRevB.84.054424} {\bibfield
  {journal} {\bibinfo  {journal} {Phys. Rev. B}\ }\textbf {\bibinfo {volume}
  {84}},\ \bibinfo {pages} {054424} (\bibinfo {year} {2011})}\BibitemShut
  {NoStop}%
\bibitem [{\citenamefont {Chen}\ \emph {et~al.}(1990)\citenamefont {Chen},
  \citenamefont {{De Gasperis}}, \citenamefont {Marcelli}, \citenamefont
  {Pardavi-Horvath}, \citenamefont {McMichael},\ and\ \citenamefont
  {Wigen}}]{Chen1990}%
  \BibitemOpen
  \bibfield  {author} {\bibinfo {author} {\bibfnamefont {H.}~\bibnamefont
  {Chen}}, \bibinfo {author} {\bibfnamefont {P.}~\bibnamefont {{De Gasperis}}},
  \bibinfo {author} {\bibfnamefont {R.}~\bibnamefont {Marcelli}}, \bibinfo
  {author} {\bibfnamefont {M.}~\bibnamefont {Pardavi-Horvath}}, \bibinfo
  {author} {\bibfnamefont {R.}~\bibnamefont {McMichael}},\ and\ \bibinfo
  {author} {\bibfnamefont {P.~E.}\ \bibnamefont {Wigen}},\ }\bibfield  {title}
  {\bibinfo {title} {{Wide-band linewidth measurements in yttrium iron garnet
  films}},\ }\href {https://doi.org/10.1063/1.345874} {\bibfield  {journal}
  {\bibinfo  {journal} {Journ. Appl. Phys.}\ }\textbf {\bibinfo {volume}
  {67}},\ \bibinfo {pages} {5530} (\bibinfo {year} {1990})}\BibitemShut
  {NoStop}%
\bibitem [{\citenamefont {Woltersdorf}\ \emph {et~al.}(2009)\citenamefont
  {Woltersdorf}, \citenamefont {Kiessling}, \citenamefont {Meyer},
  \citenamefont {Thiele},\ and\ \citenamefont {Back}}]{Woltersdorf2009}%
  \BibitemOpen
  \bibfield  {author} {\bibinfo {author} {\bibfnamefont {G.}~\bibnamefont
  {Woltersdorf}}, \bibinfo {author} {\bibfnamefont {M.}~\bibnamefont
  {Kiessling}}, \bibinfo {author} {\bibfnamefont {G.}~\bibnamefont {Meyer}},
  \bibinfo {author} {\bibfnamefont {J.-U.}\ \bibnamefont {Thiele}},\ and\
  \bibinfo {author} {\bibfnamefont {C.~H.}\ \bibnamefont {Back}},\ }\bibfield
  {title} {\bibinfo {title} {{Damping by Slow Relaxing Rare Earth Impurities in
  \ch{Ni80Fe20}}},\ }\href {https://doi.org/10.1103/PhysRevLett.102.257602}
  {\bibfield  {journal} {\bibinfo  {journal} {Phys. Rev. Lett.}\ }\textbf
  {\bibinfo {volume} {102}},\ \bibinfo {pages} {257602} (\bibinfo {year}
  {2009})}\BibitemShut {NoStop}%
\bibitem [{\citenamefont {J{\o}rgensen}\ \emph {et~al.}(2007)\citenamefont
  {J{\o}rgensen}, \citenamefont {Mosegaard}, \citenamefont {Thomsen},
  \citenamefont {Jensen},\ and\ \citenamefont {Hanson}}]{Jorgensen2007}%
  \BibitemOpen
  \bibfield  {author} {\bibinfo {author} {\bibfnamefont {J.-E.}\ \bibnamefont
  {J{\o}rgensen}}, \bibinfo {author} {\bibfnamefont {L.}~\bibnamefont
  {Mosegaard}}, \bibinfo {author} {\bibfnamefont {L.~E.}\ \bibnamefont
  {Thomsen}}, \bibinfo {author} {\bibfnamefont {T.~R.}\ \bibnamefont
  {Jensen}},\ and\ \bibinfo {author} {\bibfnamefont {J.~C.}\ \bibnamefont
  {Hanson}},\ }\bibfield  {title} {\bibinfo {title} {{Formation of
  $\gamma$-\ch{Fe2O3} nanoparticles and vacancy ordering: An in situ X-ray
  powder diffraction study}},\ }\href
  {https://doi.org/10.1016/j.jssc.2006.09.033} {\bibfield  {journal} {\bibinfo
  {journal} {J. Solid State Chem.}\ }\textbf {\bibinfo {volume} {180}},\
  \bibinfo {pages} {180} (\bibinfo {year} {2007})}\BibitemShut {NoStop}%
\bibitem [{Vil()}]{Villars2016:sm_isp_sd_0310770}%
  \BibitemOpen
  \href {https://materials.springer.com/isp/crystallographic/docs/sd_0310770}
  {\bibinfo {title} {Mgo crystal structure: Datasheet from ``pauling file
  multinaries edition -- 2012'' in springermaterials}},\ \bibinfo {note}
  {copyright 2016 Springer-Verlag Berlin Heidelberg {\&} Material Phases Data
  System (MPDS), Switzerland {\&} National Institute for Materials Science
  (NIMS), Japan}\BibitemShut {NoStop}%
\bibitem [{\citenamefont {Chicot}\ \emph {et~al.}(2011)\citenamefont {Chicot},
  \citenamefont {Mendoza}, \citenamefont {Zaoui}, \citenamefont {Louis},
  \citenamefont {Lepingle}, \citenamefont {Roudet},\ and\ \citenamefont
  {Lesage}}]{Chicot2011}%
  \BibitemOpen
  \bibfield  {author} {\bibinfo {author} {\bibfnamefont {D.}~\bibnamefont
  {Chicot}}, \bibinfo {author} {\bibfnamefont {J.}~\bibnamefont {Mendoza}},
  \bibinfo {author} {\bibfnamefont {A.}~\bibnamefont {Zaoui}}, \bibinfo
  {author} {\bibfnamefont {G.}~\bibnamefont {Louis}}, \bibinfo {author}
  {\bibfnamefont {V.}~\bibnamefont {Lepingle}}, \bibinfo {author}
  {\bibfnamefont {F.}~\bibnamefont {Roudet}},\ and\ \bibinfo {author}
  {\bibfnamefont {J.}~\bibnamefont {Lesage}},\ }\bibfield  {title} {\bibinfo
  {title} {{Mechanical properties of magnetite (\ch{Fe3O4}), hematite
  ($\alpha$-\ch{Fe2O3}) and goethite ($\alpha$-FeO{\textperiodcentered}OH) by
  instrumented indentation and molecular dynamics analysis}},\ }\href
  {https://doi.org/10.1016/j.matchemphys.2011.05.056} {\bibfield  {journal}
  {\bibinfo  {journal} {Mater. Chem. Phys.}\ }\textbf {\bibinfo {volume}
  {129}},\ \bibinfo {pages} {862} (\bibinfo {year} {2011})}\BibitemShut
  {NoStop}%
\bibitem [{\citenamefont {van~der Pauw}(1991)}]{VanderPAUW1991a}%
  \BibitemOpen
  \bibfield  {author} {\bibinfo {author} {\bibfnamefont {L.~J.}\ \bibnamefont
  {van~der Pauw}},\ }\bibfield  {title} {\bibinfo {title} {{A method of
  measuring specific resistivity and Hall effect of discs of arbitrary
  shape}},\ }\href {https://doi.org/10.1142/9789814503464_0017} {\bibfield
  {journal} {\bibinfo  {journal} {Semicond. Devices: Pion. Papers}\ ,\ \bibinfo
  {pages} {174}} (\bibinfo {year} {1991})}\BibitemShut {NoStop}%
\bibitem [{\citenamefont {Morin}(1950)}]{Morin1950}%
  \BibitemOpen
  \bibfield  {author} {\bibinfo {author} {\bibfnamefont {F.~J.}\ \bibnamefont
  {Morin}},\ }\bibfield  {title} {\bibinfo {title} {{Magnetic Susceptibility of
  $\alpha$-\ch{Fe2O3} and $\alpha$-\ch{Fe2O3} with Added Titanium}},\ }\href
  {https://doi.org/10.1103/PhysRev.78.819.2} {\bibfield  {journal} {\bibinfo
  {journal} {Phys. Rev.}\ }\textbf {\bibinfo {volume} {78}},\ \bibinfo {pages}
  {819} (\bibinfo {year} {1950})}\BibitemShut {NoStop}%
\bibitem [{\citenamefont {Verwey}(1939)}]{VERWEY1939}%
  \BibitemOpen
  \bibfield  {author} {\bibinfo {author} {\bibfnamefont {E.~J.~W.}\
  \bibnamefont {Verwey}},\ }\bibfield  {title} {\bibinfo {title} {{Electronic
  Conduction of Magnetite (\ch{Fe3O4}) and its Transition Point at Low
  Temperatures}},\ }\href {https://doi.org/10.1038/144327b0} {\bibfield
  {journal} {\bibinfo  {journal} {Nature}\ }\textbf {\bibinfo {volume} {144}},\
  \bibinfo {pages} {327} (\bibinfo {year} {1939})}\BibitemShut {NoStop}%
\bibitem [{\citenamefont {Popova}\ \emph {et~al.}(2001)\citenamefont {Popova},
  \citenamefont {Keller}, \citenamefont {Gendron}, \citenamefont {Thomas},
  \citenamefont {Brianso}, \citenamefont {Guyot}, \citenamefont {Tessier},\
  and\ \citenamefont {Parkin}}]{Popova2001}%
  \BibitemOpen
  \bibfield  {author} {\bibinfo {author} {\bibfnamefont {E.}~\bibnamefont
  {Popova}}, \bibinfo {author} {\bibfnamefont {N.}~\bibnamefont {Keller}},
  \bibinfo {author} {\bibfnamefont {F.}~\bibnamefont {Gendron}}, \bibinfo
  {author} {\bibfnamefont {L.}~\bibnamefont {Thomas}}, \bibinfo {author}
  {\bibfnamefont {M.-C.}\ \bibnamefont {Brianso}}, \bibinfo {author}
  {\bibfnamefont {M.}~\bibnamefont {Guyot}}, \bibinfo {author} {\bibfnamefont
  {M.}~\bibnamefont {Tessier}},\ and\ \bibinfo {author} {\bibfnamefont
  {S.~S.~P.}\ \bibnamefont {Parkin}},\ }\bibfield  {title} {\bibinfo {title}
  {{Perpendicular magnetic anisotropy in ultrathin yttrium iron garnet films
  prepared by pulsed laser deposition technique}},\ }\href
  {https://doi.org/10.1116/1.1392395} {\bibfield  {journal} {\bibinfo
  {journal} {J. Vac. Sci. Technol. A}\ }\textbf {\bibinfo {volume} {19}},\
  \bibinfo {pages} {2567} (\bibinfo {year} {2001})}\BibitemShut {NoStop}%
\bibitem [{\citenamefont {Bobo}\ \emph {et~al.}(2001)\citenamefont {Bobo},
  \citenamefont {Basso}, \citenamefont {Snoeck}, \citenamefont {Gatel},
  \citenamefont {Hrabovsky}, \citenamefont {Gauffier}, \citenamefont {Ressier},
  \citenamefont {Mamy}, \citenamefont {Visnovsky}, \citenamefont {Hamrle},
  \citenamefont {Teillet},\ and\ \citenamefont {Fert}}]{Bobo2001}%
  \BibitemOpen
  \bibfield  {author} {\bibinfo {author} {\bibfnamefont {J.}~\bibnamefont
  {Bobo}}, \bibinfo {author} {\bibfnamefont {D.}~\bibnamefont {Basso}},
  \bibinfo {author} {\bibfnamefont {E.}~\bibnamefont {Snoeck}}, \bibinfo
  {author} {\bibfnamefont {C.}~\bibnamefont {Gatel}}, \bibinfo {author}
  {\bibfnamefont {D.}~\bibnamefont {Hrabovsky}}, \bibinfo {author}
  {\bibfnamefont {J.}~\bibnamefont {Gauffier}}, \bibinfo {author}
  {\bibfnamefont {L.}~\bibnamefont {Ressier}}, \bibinfo {author} {\bibfnamefont
  {R.}~\bibnamefont {Mamy}}, \bibinfo {author} {\bibfnamefont {S.}~\bibnamefont
  {Visnovsky}}, \bibinfo {author} {\bibfnamefont {J.}~\bibnamefont {Hamrle}},
  \bibinfo {author} {\bibfnamefont {J.}~\bibnamefont {Teillet}},\ and\ \bibinfo
  {author} {\bibfnamefont {A.}~\bibnamefont {Fert}},\ }\bibfield  {title}
  {\bibinfo {title} {{Magnetic behavior and role of the antiphase boundaries in
  \ch{Fe3O4} epitaxial films sputtered on MgO (001)}},\ }\href
  {https://doi.org/10.1007/s100510170020} {\bibfield  {journal} {\bibinfo
  {journal} {Eur. Phys. J. B}\ }\textbf {\bibinfo {volume} {24}},\ \bibinfo
  {pages} {43} (\bibinfo {year} {2001})}\BibitemShut {NoStop}%
\bibitem [{\citenamefont {Singh}\ \emph {et~al.}(2017)\citenamefont {Singh},
  \citenamefont {Khodadadi}, \citenamefont {Mohammadi}, \citenamefont
  {Keshavarz}, \citenamefont {Mewes}, \citenamefont {Negi}, \citenamefont
  {Datta}, \citenamefont {Galazka}, \citenamefont {Uecker},\ and\ \citenamefont
  {Gupta}}]{Singh2017}%
  \BibitemOpen
  \bibfield  {author} {\bibinfo {author} {\bibfnamefont {A.~V.}\ \bibnamefont
  {Singh}}, \bibinfo {author} {\bibfnamefont {B.}~\bibnamefont {Khodadadi}},
  \bibinfo {author} {\bibfnamefont {J.~B.}\ \bibnamefont {Mohammadi}}, \bibinfo
  {author} {\bibfnamefont {S.}~\bibnamefont {Keshavarz}}, \bibinfo {author}
  {\bibfnamefont {T.}~\bibnamefont {Mewes}}, \bibinfo {author} {\bibfnamefont
  {D.~S.}\ \bibnamefont {Negi}}, \bibinfo {author} {\bibfnamefont
  {R.}~\bibnamefont {Datta}}, \bibinfo {author} {\bibfnamefont
  {Z.}~\bibnamefont {Galazka}}, \bibinfo {author} {\bibfnamefont
  {R.}~\bibnamefont {Uecker}},\ and\ \bibinfo {author} {\bibfnamefont
  {A.}~\bibnamefont {Gupta}},\ }\bibfield  {title} {\bibinfo {title} {{Bulk
  Single Crystal‐Like Structural and Magnetic Characteristics of Epitaxial
  Spinel Ferrite Thin Films with Elimination of Antiphase Boundaries}},\ }\href
  {https://doi.org/10.1002/adma.201701222} {\bibfield  {journal} {\bibinfo
  {journal} {Adv. Mater.}\ }\textbf {\bibinfo {volume} {29}},\ \bibinfo {pages}
  {1701222} (\bibinfo {year} {2017})}\BibitemShut {NoStop}%
\bibitem [{\citenamefont {Lubitz}\ \emph {et~al.}(2001)\citenamefont {Lubitz},
  \citenamefont {Rubinstein}, \citenamefont {Krebs},\ and\ \citenamefont
  {Cheng}}]{Lubitz2001}%
  \BibitemOpen
  \bibfield  {author} {\bibinfo {author} {\bibfnamefont {P.}~\bibnamefont
  {Lubitz}}, \bibinfo {author} {\bibfnamefont {M.}~\bibnamefont {Rubinstein}},
  \bibinfo {author} {\bibfnamefont {J.~J.}\ \bibnamefont {Krebs}},\ and\
  \bibinfo {author} {\bibfnamefont {S.-F.}\ \bibnamefont {Cheng}},\ }\bibfield
  {title} {\bibinfo {title} {{Frequency and temperature dependence of
  ferromagnetic linewidth in exchange biased Permalloy}},\ }\href
  {https://doi.org/10.1063/1.1358824} {\bibfield  {journal} {\bibinfo
  {journal} {Journ. Appl. Phys.}\ }\textbf {\bibinfo {volume} {89}},\ \bibinfo
  {pages} {6901} (\bibinfo {year} {2001})}\BibitemShut {NoStop}%
\bibitem [{\citenamefont {McMichael}\ \emph {et~al.}(2000)\citenamefont
  {McMichael}, \citenamefont {Lee}, \citenamefont {Stiles}, \citenamefont
  {Serpa}, \citenamefont {Chen},\ and\ \citenamefont
  {Egelhoff}}]{McMichael2000}%
  \BibitemOpen
  \bibfield  {author} {\bibinfo {author} {\bibfnamefont {R.~D.}\ \bibnamefont
  {McMichael}}, \bibinfo {author} {\bibfnamefont {C.~G.}\ \bibnamefont {Lee}},
  \bibinfo {author} {\bibfnamefont {M.~D.}\ \bibnamefont {Stiles}}, \bibinfo
  {author} {\bibfnamefont {F.~G.}\ \bibnamefont {Serpa}}, \bibinfo {author}
  {\bibfnamefont {P.~J.}\ \bibnamefont {Chen}},\ and\ \bibinfo {author}
  {\bibfnamefont {W.~F.}\ \bibnamefont {Egelhoff}},\ }\bibfield  {title}
  {\bibinfo {title} {{Exchange bias relaxation in CoO-biased films}},\ }\href
  {https://doi.org/10.1063/1.373424} {\bibfield  {journal} {\bibinfo  {journal}
  {Journ. Appl. Phys.}\ }\textbf {\bibinfo {volume} {87}},\ \bibinfo {pages}
  {6406} (\bibinfo {year} {2000})}\BibitemShut {NoStop}%
\bibitem [{\citenamefont {Knittel}\ \emph {et~al.}(2006)\citenamefont
  {Knittel}, \citenamefont {Wei}, \citenamefont {Zhou}, \citenamefont {Arora},
  \citenamefont {Shvets}, \citenamefont {Luysberg},\ and\ \citenamefont
  {Hartmann}}]{Knittel2006}%
  \BibitemOpen
  \bibfield  {author} {\bibinfo {author} {\bibfnamefont {I.}~\bibnamefont
  {Knittel}}, \bibinfo {author} {\bibfnamefont {J.}~\bibnamefont {Wei}},
  \bibinfo {author} {\bibfnamefont {Y.}~\bibnamefont {Zhou}}, \bibinfo {author}
  {\bibfnamefont {S.~K.}\ \bibnamefont {Arora}}, \bibinfo {author}
  {\bibfnamefont {I.~V.}\ \bibnamefont {Shvets}}, \bibinfo {author}
  {\bibfnamefont {M.}~\bibnamefont {Luysberg}},\ and\ \bibinfo {author}
  {\bibfnamefont {U.}~\bibnamefont {Hartmann}},\ }\bibfield  {title} {\bibinfo
  {title} {{Observation of antiferromagnetic coupling in epitaxial ferrite
  films}},\ }\href {https://doi.org/10.1103/PhysRevB.74.132406} {\bibfield
  {journal} {\bibinfo  {journal} {Phys. Rev. B}\ }\textbf {\bibinfo {volume}
  {74}},\ \bibinfo {pages} {132406} (\bibinfo {year} {2006})}\BibitemShut
  {NoStop}%
\bibitem [{\citenamefont {Hurben}\ and\ \citenamefont
  {Patton}(1998)}]{Hurben1998}%
  \BibitemOpen
  \bibfield  {author} {\bibinfo {author} {\bibfnamefont {M.~J.}\ \bibnamefont
  {Hurben}}\ and\ \bibinfo {author} {\bibfnamefont {C.~E.}\ \bibnamefont
  {Patton}},\ }\bibfield  {title} {\bibinfo {title} {{Theory of two magnon
  scattering microwave relaxation and ferromagnetic resonance linewidth in
  magnetic thin films}},\ }\href {https://doi.org/10.1063/1.367194} {\bibfield
  {journal} {\bibinfo  {journal} {Journ. Appl. Phys.}\ }\textbf {\bibinfo
  {volume} {83}},\ \bibinfo {pages} {4344} (\bibinfo {year}
  {1998})}\BibitemShut {NoStop}%
\bibitem [{\citenamefont {Polder}(1949)}]{Polder1949}%
  \BibitemOpen
  \bibfield  {author} {\bibinfo {author} {\bibfnamefont {D.}~\bibnamefont
  {Polder}},\ }\bibfield  {title} {\bibinfo {title} {{On the theory of
  ferromagnetic resonance}},\ }\href
  {https://doi.org/10.1016/0031-8914(49)90051-8} {\bibfield  {journal}
  {\bibinfo  {journal} {Physica}\ }\textbf {\bibinfo {volume} {15}},\ \bibinfo
  {pages} {253} (\bibinfo {year} {1949})}\BibitemShut {NoStop}%
\bibitem [{\citenamefont {Shaw}\ \emph {et~al.}(2014)\citenamefont {Shaw},
  \citenamefont {Nembach},\ and\ \citenamefont {Silva}}]{Shaw2014}%
  \BibitemOpen
  \bibfield  {author} {\bibinfo {author} {\bibfnamefont {J.~M.}\ \bibnamefont
  {Shaw}}, \bibinfo {author} {\bibfnamefont {H.~T.}\ \bibnamefont {Nembach}},\
  and\ \bibinfo {author} {\bibfnamefont {T.~J.}\ \bibnamefont {Silva}},\
  }\bibfield  {title} {{\selectlanguage {English}\bibinfo {title} {{Resolving
  the controversy of a possible relationship between perpendicular magnetic
  anisotropy and the magnetic damping parameter}}},\ }\href
  {https://doi.org/10.1063/1.4892532} {\bibfield  {journal} {\bibinfo
  {journal} {Appl. Phys. Lett.}\ }\textbf {\bibinfo {volume} {105}},\ \bibinfo
  {pages} {062406} (\bibinfo {year} {2014})}\BibitemShut {NoStop}%
\bibitem [{\citenamefont {Youssef}\ and\ \citenamefont
  {Brosseau}(2006)}]{Youssef2006}%
  \BibitemOpen
  \bibfield  {author} {\bibinfo {author} {\bibfnamefont {J.~B.}\ \bibnamefont
  {Youssef}}\ and\ \bibinfo {author} {\bibfnamefont {C.}~\bibnamefont
  {Brosseau}},\ }\bibfield  {title} {\bibinfo {title} {{Magnetization damping
  in two-component metal oxide micropowder and nanopowder compacts by broadband
  ferromagnetic resonance measurements}},\ }\href
  {https://doi.org/10.1103/PhysRevB.74.214413} {\bibfield  {journal} {\bibinfo
  {journal} {Phys. Rev. B}\ }\textbf {\bibinfo {volume} {74}},\ \bibinfo
  {pages} {214413} (\bibinfo {year} {2006})}\BibitemShut {NoStop}%
\bibitem [{\citenamefont {Takeda}\ \emph {et~al.}(2009)\citenamefont {Takeda},
  \citenamefont {Onishi}, \citenamefont {Nakakubo},\ and\ \citenamefont
  {Fujimoto}}]{Takeda2009}%
  \BibitemOpen
  \bibfield  {author} {\bibinfo {author} {\bibfnamefont {M.}~\bibnamefont
  {Takeda}}, \bibinfo {author} {\bibfnamefont {T.}~\bibnamefont {Onishi}},
  \bibinfo {author} {\bibfnamefont {S.}~\bibnamefont {Nakakubo}},\ and\
  \bibinfo {author} {\bibfnamefont {S.}~\bibnamefont {Fujimoto}},\ }\bibfield
  {title} {\bibinfo {title} {{Physical properties of iron-oxide scales on
  Si-containing steels at high temperature}},\ }\href
  {https://doi.org/10.2320/matertrans.M2009097} {\bibfield  {journal} {\bibinfo
   {journal} {Mater. Trans.}\ }\textbf {\bibinfo {volume} {50}},\ \bibinfo
  {pages} {2242} (\bibinfo {year} {2009})}\BibitemShut {NoStop}%
\bibitem [{\citenamefont {Madelung}\ \emph {et~al.}(1999)\citenamefont
  {Madelung}, \citenamefont {R{\"{o}}ssler},\ and\ \citenamefont
  {Schulz}}]{Madelung1999}%
  \BibitemOpen
  \bibinfo {editor} {\bibfnamefont {O.}~\bibnamefont {Madelung}}, \bibinfo
  {editor} {\bibfnamefont {U.}~\bibnamefont {R{\"{o}}ssler}},\ and\ \bibinfo
  {editor} {\bibfnamefont {M.}~\bibnamefont {Schulz}},\ eds.,\ \href
  {https://doi.org/10.1007/b71137} {\emph {\bibinfo {title} {{II-VI and I-VII
  Compounds; Semimagnetic Compounds}}}},\ \bibinfo {series}
  {Landolt-B{\"{o}}rnstein - Group III Condensed Matter}, Vol.\ \bibinfo
  {volume} {41B}\ (\bibinfo  {publisher} {Springer-Verlag},\ \bibinfo {address}
  {Berlin/Heidelberg},\ \bibinfo {year} {1999})\BibitemShut {NoStop}%
\bibitem [{\citenamefont {Goodenough}\ \emph {et~al.}(1970)\citenamefont
  {Goodenough}, \citenamefont {Gräper}, \citenamefont {Holtzberg},
  \citenamefont {Huber}, \citenamefont {Lefever}, \citenamefont {Longo},
  \citenamefont {McGuire},\ and\ \citenamefont {Methfessel}}]{Overview2015}%
  \BibitemOpen
  \bibfield  {author} {\bibinfo {author} {\bibfnamefont {J.}~\bibnamefont
  {Goodenough}}, \bibinfo {author} {\bibfnamefont {W.}~\bibnamefont {Gräper}},
  \bibinfo {author} {\bibfnamefont {F.}~\bibnamefont {Holtzberg}}, \bibinfo
  {author} {\bibfnamefont {D.}~\bibnamefont {Huber}}, \bibinfo {author}
  {\bibfnamefont {R.}~\bibnamefont {Lefever}}, \bibinfo {author} {\bibfnamefont
  {J.}~\bibnamefont {Longo}}, \bibinfo {author} {\bibfnamefont
  {T.}~\bibnamefont {McGuire}},\ and\ \bibinfo {author} {\bibfnamefont
  {S.}~\bibnamefont {Methfessel}},\ }\href {https://doi.org/10.1007/b19968}
  {\emph {\bibinfo {title} {{Part A}}}},\ edited by\ \bibinfo {editor}
  {\bibfnamefont {K.-H.}\ \bibnamefont {Hellwege}}\ and\ \bibinfo {editor}
  {\bibfnamefont {A.~M.}\ \bibnamefont {Hellwege}},\ \bibinfo {series}
  {Landolt-B{\"{o}}rnstein - Group III Condensed Matter}, Vol.~\bibinfo
  {volume} {4a}\ (\bibinfo  {publisher} {Springer-Verlag},\ \bibinfo {address}
  {Berlin/Heidelberg},\ \bibinfo {year} {1970})\ pp.\ \bibinfo {pages}
  {1--2}\BibitemShut {NoStop}%
\bibitem [{\citenamefont {Drovosekov}\ \emph {et~al.}(2020)\citenamefont
  {Drovosekov}, \citenamefont {Kreines}, \citenamefont {Barkalova},
  \citenamefont {Nikolaev}, \citenamefont {Sitnikov},\ and\ \citenamefont
  {Rylkov}}]{Drovosekov2020}%
  \BibitemOpen
  \bibfield  {author} {\bibinfo {author} {\bibfnamefont {A.~B.}\ \bibnamefont
  {Drovosekov}}, \bibinfo {author} {\bibfnamefont {N.~M.}\ \bibnamefont
  {Kreines}}, \bibinfo {author} {\bibfnamefont {A.~S.}\ \bibnamefont
  {Barkalova}}, \bibinfo {author} {\bibfnamefont {S.~N.}\ \bibnamefont
  {Nikolaev}}, \bibinfo {author} {\bibfnamefont {A.~V.}\ \bibnamefont
  {Sitnikov}},\ and\ \bibinfo {author} {\bibfnamefont {V.~V.}\ \bibnamefont
  {Rylkov}},\ }\bibfield  {title} {\bibinfo {title} {{Effect of Slow Ion
  Relaxation at Ferromagnetic Resonance in a CoFeB-LiNbO Metal-Insulator
  Nanocomposite}},\ }\href {https://doi.org/10.1134/S0021364020140088}
  {\bibfield  {journal} {\bibinfo  {journal} {JETP Lett.}\ }\textbf {\bibinfo
  {volume} {112}},\ \bibinfo {pages} {84} (\bibinfo {year} {2020})}\BibitemShut
  {NoStop}%
\bibitem [{\citenamefont {Clarke}\ \emph {et~al.}(1965)\citenamefont {Clarke},
  \citenamefont {Tweedale},\ and\ \citenamefont {Teale}}]{Bleaney1961}%
  \BibitemOpen
  \bibfield  {author} {\bibinfo {author} {\bibfnamefont {B.~H.}\ \bibnamefont
  {Clarke}}, \bibinfo {author} {\bibfnamefont {K.}~\bibnamefont {Tweedale}},\
  and\ \bibinfo {author} {\bibfnamefont {R.~W.}\ \bibnamefont {Teale}},\
  }\bibfield  {title} {\bibinfo {title} {{Rare-Earth Ion Relaxation Time and
  $G$ Tensor in Rare-Earth-Doped Yttrium Iron Garnet. I. Ytterbium}},\ }\href
  {https://doi.org/10.1103/PhysRev.139.A1933} {\bibfield  {journal} {\bibinfo
  {journal} {Phys. Rev.}\ }\textbf {\bibinfo {volume} {139}},\ \bibinfo {pages}
  {A1933} (\bibinfo {year} {1965})}\BibitemShut {NoStop}%
\bibitem [{\citenamefont {Orbach}(1961)}]{Rfsoyay1961}%
  \BibitemOpen
  \bibfield  {author} {\bibinfo {author} {\bibfnamefont {R.}~\bibnamefont
  {Orbach}},\ }\bibfield  {title} {\bibinfo {title} {{Spin-lattice relaxation
  in rare-earth salts}},\ }\href {https://doi.org/10.1098/rspa.1961.0211}
  {\bibfield  {journal} {\bibinfo  {journal} {Proc. R. Soc. A: Math. Phys. Eng.
  Sci.}\ }\textbf {\bibinfo {volume} {264}},\ \bibinfo {pages} {458} (\bibinfo
  {year} {1961})}\BibitemShut {NoStop}%
\bibitem [{\citenamefont {Hansen}\ \emph {et~al.}(1973)\citenamefont {Hansen},
  \citenamefont {Schuldt},\ and\ \citenamefont {Tolksdorf}}]{Hansen1973}%
  \BibitemOpen
  \bibfield  {author} {\bibinfo {author} {\bibfnamefont {P.}~\bibnamefont
  {Hansen}}, \bibinfo {author} {\bibfnamefont {J.}~\bibnamefont {Schuldt}},\
  and\ \bibinfo {author} {\bibfnamefont {W.}~\bibnamefont {Tolksdorf}},\
  }\bibfield  {title} {\bibinfo {title} {{Anisotropy and Magnetostriction of
  Iridium-Substituted Yttrium Iron Garnet}},\ }\href
  {https://doi.org/10.1103/PhysRevB.8.4274} {\bibfield  {journal} {\bibinfo
  {journal} {Phys. Rev. B}\ }\textbf {\bibinfo {volume} {8}},\ \bibinfo {pages}
  {4274} (\bibinfo {year} {1973})}\BibitemShut {NoStop}%
\bibitem [{\citenamefont {Hartwick}\ and\ \citenamefont
  {Smit}(1968)}]{Hartwick1968}%
  \BibitemOpen
  \bibfield  {author} {\bibinfo {author} {\bibfnamefont {T.~S.}\ \bibnamefont
  {Hartwick}}\ and\ \bibinfo {author} {\bibfnamefont {J.}~\bibnamefont
  {Smit}},\ }\bibfield  {title} {\bibinfo {title} {{Ferromagnetic Resonance in
  Si‐Doped YIG}},\ }\href {https://doi.org/10.1063/1.2163631} {\bibfield
  {journal} {\bibinfo  {journal} {Journ. Appl. Phys.}\ }\textbf {\bibinfo
  {volume} {39}},\ \bibinfo {pages} {827} (\bibinfo {year} {1968})}\BibitemShut
  {NoStop}%
\bibitem [{\citenamefont {Spencer}\ \emph {et~al.}(1961)\citenamefont
  {Spencer}, \citenamefont {LeCraw},\ and\ \citenamefont
  {Linares}}]{Spencer1961}%
  \BibitemOpen
  \bibfield  {author} {\bibinfo {author} {\bibfnamefont {E.~G.}\ \bibnamefont
  {Spencer}}, \bibinfo {author} {\bibfnamefont {R.~C.}\ \bibnamefont
  {LeCraw}},\ and\ \bibinfo {author} {\bibfnamefont {R.~C.}\ \bibnamefont
  {Linares}},\ }\bibfield  {title} {\bibinfo {title} {{Low-Temperature
  Ferromagnetic Relaxation in Yttrium Iron Garnet}},\ }\href
  {https://doi.org/10.1103/PhysRev.123.1937} {\bibfield  {journal} {\bibinfo
  {journal} {Phys. Rev.}\ }\textbf {\bibinfo {volume} {123}},\ \bibinfo {pages}
  {1937} (\bibinfo {year} {1961})}\BibitemShut {NoStop}%
\bibitem [{\citenamefont {Hansen}\ \emph {et~al.}(1972)\citenamefont {Hansen},
  \citenamefont {Tolksdorf},\ and\ \citenamefont {Schuldt}}]{Hansen1973a}%
  \BibitemOpen
  \bibfield  {author} {\bibinfo {author} {\bibfnamefont {P.}~\bibnamefont
  {Hansen}}, \bibinfo {author} {\bibfnamefont {W.}~\bibnamefont {Tolksdorf}},\
  and\ \bibinfo {author} {\bibfnamefont {J.}~\bibnamefont {Schuldt}},\
  }\bibfield  {title} {\bibinfo {title} {{Anisotropy and magnetostriction of
  germanium‐substituted yttrium iron garnet}},\ }\href
  {https://doi.org/10.1063/1.1660999} {\bibfield  {journal} {\bibinfo
  {journal} {Journ. Appl. Phys.}\ }\textbf {\bibinfo {volume} {43}},\ \bibinfo
  {pages} {4740} (\bibinfo {year} {1972})}\BibitemShut {NoStop}%
\bibitem [{\citenamefont {Epstein}\ and\ \citenamefont
  {Tocci}(1967)}]{Epstein1967}%
  \BibitemOpen
  \bibfield  {author} {\bibinfo {author} {\bibfnamefont {D.~J.}\ \bibnamefont
  {Epstein}}\ and\ \bibinfo {author} {\bibfnamefont {L.}~\bibnamefont
  {Tocci}},\ }\bibfield  {title} {\bibinfo {title} {{High temperature resonance
  losses in silicon-doped Yttrium-iron garnet (YIG)}},\ }\href
  {https://doi.org/10.1063/1.1755026} {\bibfield  {journal} {\bibinfo
  {journal} {Appl. Phys. Lett.}\ }\textbf {\bibinfo {volume} {11}},\ \bibinfo
  {pages} {55} (\bibinfo {year} {1967})}\BibitemShut {NoStop}%
\bibitem [{\citenamefont {Coduri}\ \emph {et~al.}(2020)\citenamefont {Coduri},
  \citenamefont {Masala}, \citenamefont {{Del Bianco}}, \citenamefont {Spizzo},
  \citenamefont {Ceresoli}, \citenamefont {Castellano}, \citenamefont
  {Cappelli}, \citenamefont {Oliva}, \citenamefont {Checchia}, \citenamefont
  {Allieta}, \citenamefont {Szabo}, \citenamefont {Schlabach}, \citenamefont
  {Hagelstein}, \citenamefont {Ferrero},\ and\ \citenamefont
  {Scavini}}]{Coduri2020}%
  \BibitemOpen
  \bibfield  {author} {\bibinfo {author} {\bibfnamefont {M.}~\bibnamefont
  {Coduri}}, \bibinfo {author} {\bibfnamefont {P.}~\bibnamefont {Masala}},
  \bibinfo {author} {\bibfnamefont {L.}~\bibnamefont {{Del Bianco}}}, \bibinfo
  {author} {\bibfnamefont {F.}~\bibnamefont {Spizzo}}, \bibinfo {author}
  {\bibfnamefont {D.}~\bibnamefont {Ceresoli}}, \bibinfo {author}
  {\bibfnamefont {C.}~\bibnamefont {Castellano}}, \bibinfo {author}
  {\bibfnamefont {S.}~\bibnamefont {Cappelli}}, \bibinfo {author}
  {\bibfnamefont {C.}~\bibnamefont {Oliva}}, \bibinfo {author} {\bibfnamefont
  {S.}~\bibnamefont {Checchia}}, \bibinfo {author} {\bibfnamefont
  {M.}~\bibnamefont {Allieta}}, \bibinfo {author} {\bibfnamefont {D.-V.}\
  \bibnamefont {Szabo}}, \bibinfo {author} {\bibfnamefont {S.}~\bibnamefont
  {Schlabach}}, \bibinfo {author} {\bibfnamefont {M.}~\bibnamefont
  {Hagelstein}}, \bibinfo {author} {\bibfnamefont {C.}~\bibnamefont
  {Ferrero}},\ and\ \bibinfo {author} {\bibfnamefont {M.}~\bibnamefont
  {Scavini}},\ }\bibfield  {title} {\bibinfo {title} {{Local Structure and
  Magnetism of \ch{Fe2O3} Maghemite Nanocrystals: The Role of Crystal
  Dimension}},\ }\href {https://doi.org/10.3390/nano10050867} {\bibfield
  {journal} {\bibinfo  {journal} {Nanomater.}\ }\textbf {\bibinfo {volume}
  {10}},\ \bibinfo {pages} {867} (\bibinfo {year} {2020})}\BibitemShut
  {NoStop}%
\bibitem [{\citenamefont {Kasuya}\ and\ \citenamefont
  {LeCraw}(1961)}]{Kasuya1961}%
  \BibitemOpen
  \bibfield  {author} {\bibinfo {author} {\bibfnamefont {T.}~\bibnamefont
  {Kasuya}}\ and\ \bibinfo {author} {\bibfnamefont {R.~C.}\ \bibnamefont
  {LeCraw}},\ }\bibfield  {title} {\bibinfo {title} {{Relaxation Mechanisms in
  Ferromagnetic Resonance}},\ }\href
  {https://doi.org/10.1103/PhysRevLett.6.223} {\bibfield  {journal} {\bibinfo
  {journal} {Phys. Rev. Lett.}\ }\textbf {\bibinfo {volume} {6}},\ \bibinfo
  {pages} {223} (\bibinfo {year} {1961})}\BibitemShut {NoStop}%
\bibitem [{\citenamefont {Sparks}(1964)}]{SparksMarshall1964Ft}%
  \BibitemOpen
  \bibfield  {author} {\bibinfo {author} {\bibfnamefont {M.}~\bibnamefont
  {Sparks}},\ }\href
  {https://www.science.org/doi/10.1126/science.148.3667.218.a} {\emph {\bibinfo
  {title} {Ferromagnetic-relaxation theory}}},\ McGraw-Hill advanced physics
  monograph series\ (\bibinfo  {publisher} {McGraw-Hill},\ \bibinfo {address}
  {New York},\ \bibinfo {year} {1964})\BibitemShut {NoStop}%
\bibitem [{\citenamefont {Gurevich}\ and\ \citenamefont
  {Melkov}(1996)}]{gurevich1996magnetization}%
  \BibitemOpen
  \bibfield  {author} {\bibinfo {author} {\bibfnamefont {A.}~\bibnamefont
  {Gurevich}}\ and\ \bibinfo {author} {\bibfnamefont {G.}~\bibnamefont
  {Melkov}},\ }\href {https://books.google.de/books?id=YgQtSvFIvFQC} {\emph
  {\bibinfo {title} {Magnetization Oscillations and Waves}}}\ (\bibinfo
  {publisher} {Taylor \& Francis},\ \bibinfo {year} {1996})\BibitemShut
  {NoStop}%
\bibitem [{\citenamefont {Krysztofik}\ \emph {et~al.}(2021)\citenamefont
  {Krysztofik}, \citenamefont {{\"{O}}zoğlu}, \citenamefont {McMichael},\ and\
  \citenamefont {Coy}}]{Krysztofik2021}%
  \BibitemOpen
  \bibfield  {author} {\bibinfo {author} {\bibfnamefont {A.}~\bibnamefont
  {Krysztofik}}, \bibinfo {author} {\bibfnamefont {S.}~\bibnamefont
  {{\"{O}}zoğlu}}, \bibinfo {author} {\bibfnamefont {R.~D.}\ \bibnamefont
  {McMichael}},\ and\ \bibinfo {author} {\bibfnamefont {E.}~\bibnamefont
  {Coy}},\ }\bibfield  {title} {\bibinfo {title} {{Effect of strain-induced
  anisotropy on magnetization dynamics in $\mathrm{Y_3Fe_5O_{12}}$ films
  recrystallized on a lattice-mismatched substrate}},\ }\href
  {https://doi.org/10.1038/s41598-021-93308-3} {\bibfield  {journal} {\bibinfo
  {journal} {Sci. Rep.}\ }\textbf {\bibinfo {volume} {11}},\ \bibinfo {pages}
  {14011} (\bibinfo {year} {2021})}\BibitemShut {NoStop}%
\bibitem [{\citenamefont {Clogston}(1955)}]{Clogston1955}%
  \BibitemOpen
  \bibfield  {author} {\bibinfo {author} {\bibfnamefont {A.~M.}\ \bibnamefont
  {Clogston}},\ }\bibfield  {title} {\bibinfo {title} {{Relaxation Phenomena in
  Ferrites}},\ }\href {https://doi.org/10.1002/j.1538-7305.1955.tb03774.x}
  {\bibfield  {journal} {\bibinfo  {journal} {Bell Syst. tech. j.}\ }\textbf
  {\bibinfo {volume} {34}},\ \bibinfo {pages} {739} (\bibinfo {year}
  {1955})}\BibitemShut {NoStop}%
\bibitem [{\citenamefont {Fert}\ \emph {et~al.}(2013)\citenamefont {Fert},
  \citenamefont {Cros},\ and\ \citenamefont {Sampaio}}]{Fert2013}%
  \BibitemOpen
  \bibfield  {author} {\bibinfo {author} {\bibfnamefont {A.}~\bibnamefont
  {Fert}}, \bibinfo {author} {\bibfnamefont {V.}~\bibnamefont {Cros}},\ and\
  \bibinfo {author} {\bibfnamefont {J.}~\bibnamefont {Sampaio}},\ }\bibfield
  {title} {\bibinfo {title} {{Skyrmions on the track}},\ }\href
  {https://doi.org/10.1038/nnano.2013.29} {\bibfield  {journal} {\bibinfo
  {journal} {Nat. Nanotechnol.}\ }\textbf {\bibinfo {volume} {8}},\ \bibinfo
  {pages} {152} (\bibinfo {year} {2013})}\BibitemShut {NoStop}%
\end{thebibliography}%

\end{document}